\documentclass[epj]{svjour}
\usepackage{graphics}
\usepackage{amsmath,amssymb}

\def\be{\begin{equation}}
\def\ee{\end{equation}}
\def\ba{\begin{array}}
\def\ea{\end{array}}
\def\bea{\begin{eqnarray}}
\def\eea{\end{eqnarray}}
\def\bi{\begin{itemize}}
\def\ei{\end{itemize}}
\begin{document}
\title{Ensembles of unified crust and core equations of state in a nuclear-multimessenger astrophysics environment}
\author{W.G.Newton\inst{1}, L. Balliet, S.Budimir, G. Crocombe, B. Douglas, T. Head, Z. Langford, L. Rivera \and J. Sanford\inst{1}
\thanks{william.newton@tamuc.edu, Department of Physics and Astronomy, Texas A\&M University-Commerce, Commerce, TX, 75429, USA}%
}                     
%
%
\institute{Department of Physics and Astronomy, Texas A\&M University-Commerce, Commerce, TX, 75429, USA}
\date{Received: date / Revised version: date}
%
\abstract{
We present an ensemble of unified neutron star crust and core equations of state, constructed using an extended Skyrme energy density functional through the crust and outer core, and appended by two piecewise polytropes at higher densities. The equations of state are parameterized by the first three coefficients in the density expansion of the symmetry energy $J,L$ and $K_{\rm sym}$, the moment of inertia of a 1.338 M$_{\odot}$ star $I_{1.338}$ and the maximum neutron star mass $M_{\rm max}$. We construct an ensemble with uniform priors on all five parameters, and then apply data filters to the ensemble to explore the effect of combining neutron skin data from PREX with astrophysical measurements of radii and tidal deformabilities from NICER and LIGO/VIRGO. Neutron skins are calculated directly using the EDFs. We demonstrate that both the nuclear data and astrophysical data play a role in constraining crust properties such as the mass, thickness and moment of inertia of the crust and the nuclear pasta layers therein, and that astrophysical data better constrains $K_{\rm sym}$ than PREX data. 
\PACS{
      {97.60.Jd}{Neutron stars}   \and {26.60.Dd}{core} \and   {26.60.Gj}{crust} \and {26.60.Kp}{equations of state} \and {26.60.-c}{nuclear matter aspects of}
      {PACS-key}{discribing text of that key}
     } 
} 
\authorrunning{Newton \emph{et al}}
\titlerunning{Unified EOSs in a nuclear multimessenger environment}
\maketitle
\section{Introduction}
\label{sec:intro}

The systematic exploration of the space of neutron star equations-of-state (EOSs) by generating large ensembles of EOSs which can be constrained by data using statistical inference began in earnest a decade ago \cite{Read:2009aa,Ozel:2009kx,Steiner:2010aa}. It has matured with the advent of gravitational wave measurements of the neutron star tidal deformability \cite{Abbott:2017aa,Abbott:2019kl} from LIGO/VIRGO and new measurements of the neutron star radius from timing of X-ray pulsars from the NICER telescope \cite{Riley:2019aa,Raaijmakers:2019sf,Miller:2019jt,Miller:2021hb,Riley:2021sp,Raaijmakers:2021hc}

Many different ways of generating EOS ensembles have been explored. One can characterize the EOS model with a set of physical parameters and explore the parameter space, using polytropic EOSs \cite{Read:2009aa,Ozel:2009kx,Steiner:2010aa,Raithel:2018ai,Greif:2020cr,Al-Mamun:2021ue}, line segments, speed-of-sound models \cite{Tews:2018fj,Annala:2020wd,Essick:2021vn}, and spectral models \cite{Lindblom:2010vf}. Alternatively, non-parametric EOSs can be employed. These can be generated from Gaussian processes (GPs) \cite{Landry:2019te,Landry:2020wv,Essick:2020uy,Legred:2021wn} or using machine learning techniques \cite{Ferreira:2019wh}. It should be noted that when non-parametric EOSs are used, connections to the physics of the EOS requires an extra modeling step \cite{Essick:2021vn}.

At the same time, there have been significant developments on the nuclear experimental and theoretical side: for example, measurements of the parity violating asymmetry in electron scattering off $^{208}$Pb in the PREX and CREX experiments \cite{Abrahamyan:2012ai,Becker:2018aa,Thiel:2019aa,Adhikari:2021tm}, measurements of the dipole polarizability of $^{48}$Ca and $^{208}$Pb \cite{Tamii:2011lr,Hashimoto:2015fj,von-Neumann-Cosel:2015mz,Birkhan:2017fk}, and modeling the pure neutron matter EOS using chiral effective field theory ($\chi$EFT) with well quantified theoretical uncertainty \cite{Gandolfi:2015nr,Hagen:2016ul,Tews:2016ty,Drischler:2020aa,Drischler:2020ab}. 

Including nuclear data in EOS inference typically involves use of a nuclear EOS up to around 1.5$n_{0}$ ($n_{0}$=0.16 fm$^{-3}$) that is a direct parameterization of the nuclear matter EOS expanded about nuclear saturation density, and are often fit to nuclear theory calculations of neutron matter \cite{Hebeler:2010rw,Gandolfi:2010jk,Pang:2021gf} and, more recently, heavy ion collision constraints on elliptic flow \cite{Huth:2021wr,Huth:2021lr}.

It is convenient to parameterize EOS models using the coefficients in the density expansion of the nuclear symmetry energy about nuclear saturation density $\rho_0$: defining $\chi = {(\rho - \rho_0) / 3\rho_0}$,

\vspace{-0.1cm}
{\small
\begin{equation}
S(\rho) = J + \chi L + {1 \over 2\!} \chi^2 K_{\rm sym} + {1 \over 3\!} \chi^3 Q_{\rm sym} + \dots
\end{equation}
}
\vspace{-0.3cm}

\noindent where, given the energy of nuclear matter as a function of density and proton fraction $E(\rho,x)$, $S(\rho) = d^2 E(\rho,x)/d\delta^2 |_{\delta=0}$ and $\delta = 1-2x$. The symmetry energy parameters are known to correlate with a number of neutron star properties including the proton fraction in neutron stars and the neutron star radius \cite{Brown:2000aa,Horowitz:2001aa,Steiner:2008aa,Fattoyev:2018aa}.

The full range of density dependences of nucleonic equations of state for which the predictions of EDFs form a subset can be explored in meta-models, using the parameters of the symmetric nuclear matter EOS and symmetry energy expanded to sufficiently high order \cite{Margueron:2018aa,Margueron:2018ab,Lim:2019xr,Zhang:2021ly,Biswas:2021bh}. Meta-models of this kind are inherently grounded in nucleonic physics, and assume a smooth EOS and so discount phase transitions and implicitly assume the EOS at saturation density completely determines the EOS at high densities, which might not hold \cite{Lee:2021eu}. As well as incorporating nuclear data into EOS inference, these approaches allow inference of the symmetry energy parameters from astrophysical data \cite{Tang:2021rt}, complementing the significant experimental effort that has been devoted to measuring the symmetry energy over the past two decades \cite{Tsang:2012qy,Li:2014fj,Lattimer:2014uq,Horowitz:2014aa,Tsang:2019ab,Lynch:2021qp}.

These models have started to incorporate neutron skin data using the correlation between the slope of the symmetry energy $L$ and the neutron skin thickness of $^{208}$Pb \cite{Roca-Maza:2011aa} to associate a neutron skin thickness with the nuclear matter EOS used to construct the neutron star EOS \cite{Essick:2021vn,Biswas:2021pd}. There is a fundamental inconsistency in this approach, since the correlation is obtained from set of energy density functionals (EDFs) which cover a region of symmetry energy parameter space restricted by strong correlations between $L$ and the other symmetry energy parameters from the model and the data used to fit the EDFs. Such a restriction is not enforced in the construction of the neutron star EOS, with priors that can allow independent variation of $J$, $L$, and often higher-order parameters.

Using a full energy density functional (EDF) in a Bayesian inference would allow consistent combination of finite nuclear data such as neutron skins into the multimessenger framework. Large sets of Skyrme \cite{Du:2021fk,Yue:2021ij}, Gogny \cite{Sellahewa:2014xy,Vinas:2021fv} and RMF models \cite{Dutra:2014fy,Beloin:2019oq} have been used to construct sets of neutron star EOSs, although they do not systematically explore the space of symmetry energy parameterizations, and they are usually extended to the highest densities, implicitly assuming a nucleonic star with no phase transitions. Specific parameterizations of the Skyrme EDF have been used in conjunction with polytropic and speed-of-sound parameterizations of the high density EOS \cite{Tsang:2018yq,Du:2021fk,Miao:2021fj,Li:2021kx}. Only a small number of studies have generated large ensembles of EDF models of the EOS, and additionally decoupled the high density EOS from the EDF through the use of polytropic extensions starting at 1.5-2$n_{0}$ \cite{Beloin:2019oq}. 
All ensembles of neutron star EOSs constructed with EDFs have allowed, at most, $J$ and $L$ to vary independently. However, there is significant uncertainty in the higher order parameters such as $K_{\rm sym}$ \cite{Chen:2009hh,Centelles:2009aa,Raduta:2018aa,Newton:2021dq}. There has been some limited exploration of $J,L,K_{\rm sym}$ space using a Skyrme EDF \cite{Yue:2021ij}, but a not a large systematic generation of ensembles of such models.

Explorations of the space of neutron star crust EOSs are not as developed as those of the core EOS. There exists only two attempts to systematically generate of crust model ensembles, using meta-models \cite{Carreau:2019aa,Dinh-Thi:2021lr,Lim:2019xr} and using an extended Skyrme EDF \cite{Balliet:2021lr,Newton:2021tg}.

Failure to use unified crust-core EOSs incurs an error in modeling radii and tidal deformabilities around 3\% \cite{Gamba:2020ul}, well below the current precision of observations but may be important for observations from next generation telescopes. It also decreases the accuracy of universal relations \cite{Suleiman:2021ve}. However, consistent crust and core models - unified EOSs \cite{Douchin:2001bq,Fortin:2016uq,McNeil-Forbes:2019wx,Vinas:2021fv} - gives access to a rich variety of currently observed or potential astrophysical data which measure observables including crustal oscillation modes \cite{Steiner:2009wo,Gearheart:2011tg,Sotani:2012oz}, crust cooling \cite{Brown:2009aa,Horowitz:2015rt,Newton:2013dz}, crust shattering \cite{Tsang:2012aa,Neill:2021lq}, persistent gravitational waves from mountains \cite{Gearheart:2011tg}, magnetic field evolution \cite{Pons:2013ly} and damping of core modes \cite{Wen:2012aa,Vidana:2012aa}. They allow consistent propagation of measurements of bulk neutron star properties such as radii, moments of inertia and tidal deformabilities through to crust observables.

In order to close the gaps in EOS modeling highlighted above, here we present a large ensemble of EOSs which uniquely combine the following features:

\begin{enumerate}
    \item A large number of extended Skyrme EDFs  \cite{Lim:2017aa,Balliet:2021lr} with three degrees of freedom $J$, $L$, $K_{\rm sym}$ are generated, from which neutron skin predictions are calculated.
    \item Crust models are calculated with each EDF \cite{Balliet:2021lr,Newton:2021tg}, unifying the crust and outer core EOS.
    \item Two polytropes are appended to explore a wider range of possible core EOSs.
\end{enumerate}

The polytropic parameters can be adjusted to reproduce desired values of the moment of inertia $I_{1.338}$ of a 1.338M$_{\odot}$ star - which may be measured accurately in the next few years \cite{Kramer:2021kq} - and the maximum neutron star mass $M_{\rm max}$. We construct an ensemble of EOSs with uniform priors on $J$, $L$, $K_{\rm sym}$, $I_{1.338}$, $M_{\rm max}$ which match the uniform prior distribution of crust models presented in \cite{Balliet:2021lr}. We conduct a simple initial demonstration that this set of EOSs can be used to constrain crust and core properties, including the extent of nuclear pasta in the neutron star crust, by consistently incorporating the neutron skin measurements of PREX, the neutron star tidal deformability measurements from LIGO, and the neutron star mass and radius measurements from NICER. These EOSs can in future be used directly in Bayesian inferences from combined nuclear and astrophysical datasets or, for example, as training sets for machine learning or Gaussian process generation of non-parametric EOSs.

\section{Equation of state construction}
\label{sec:eos}

We have already constructed large ensembles of extended Skyrme EDF crust models, parameterized by $J$, $L$ and $K_{\rm sym}$, using a compressible liquid drop model (CLDM) discussed throroughly in \cite{Newton:2013sp,Balliet:2021lr}. We have explored a number of prior distributions of crust models and performed a Bayesian inference of crust properties from neutron skin and neutron matter data \cite{Balliet:2021lr}.  

Here we use these models, with the best fit value of the nuclear surface parameters in the CLDM obtained by fits to 3D Hartree-Fock calculations of nuclei in the crust \cite{Balliet:2021lr}.

To systematically explore the high density equation of state in a way that is independent of the saturation-density nucleonic EOS, we employ the piecewise polytrope method \cite{Read:2009aa,Ozel:2009kx,Steiner:2010aa}. We will attach two polytropes to our Skyrme EOSs. We are essentially replacing one of the three polytropes commonly used with our extended Skyrme EOS for the outer core without losing degrees of freedom. This is comparable to recent works where the outer core EOS is described by a parameterization of the pure neutron matter EOS, usually with two degrees of freedom (equivalent of $J$ and $L$) \cite{Lim:2019xr,Capano:2020ai,Tews:2019ff}. Our model has the advantage of being derived from a full EDF that can be additionally used to calculate directly nuclear observables. The transition between the nuclear matter EOS and the first polytrope, and between the first and second polytropes, is generally found to be between 1-2$n_0$ and 2.5-4$n_0$ respectively
\cite{Read:2009aa,Steiner:2010aa,Tang:2021rt}. As our baseline family of models, we fit the first polytrope at a density of $n_1$=1.5$n_{\rm 0}$ and the second at $n_2$=2.7$n_{\rm 0}$. We then have three regions of the star: the crust and outer core, in which the pressure and energy density are given by the Skyrme EOS, and the two polytropic regions in which the pressures are given by

\begin{align}
\nonumber P_1 = K_1 n^{\Gamma_1}  & \;\;\;\;\;n_1 <n<n_2  \\
P_2 = K_2 n^{\Gamma_2} & \;\;\;\;\;  n_2 < n,
\end{align}


\noindent where continuity of pressure determines the constants $K_1$ and $K_2$. The energy density in the three density regions is obtained by integrating the first law of thermodynamics:

\be
\nonumber \epsilon_i = (1+a_i)n + \frac{K_i n^{\Gamma_i}}{\Gamma_1 -1} \;\;\;\;\;\;\; a_i = \frac{\epsilon_{i-1} (n_i)}{n_i} - \frac{K_i n^{\Gamma_i - 1}}{\Gamma_i -1} - 1
\ee

\noindent where $a_i$ are constants of integration, $i$=\{1,2\} and the subscript 0 labels the Skyrme EOS.

The speed of sound is 

\begin{equation}
\frac{c_{\rm s,i}(n)}{c} = \bigg( \frac{\Gamma_i P}{P + \epsilon}\bigg)^{1/2}
\end{equation}

In the eventuality that the EOS becomes acausal at a given density $n_{\rm acausal}$, we transition to a causal EOS:

\begin{equation}
P_{\rm causal} = \epsilon = bn^{1/3} \;\;\;\;\;  n_{\rm acausal} < n
\end{equation}

\noindent where $\epsilon$ is the energy density and $b$ is a constant given by

\begin{equation} 
b = \frac{1+a}{n_{\rm acausal}} + K \frac{n_{\rm acausal}^{\Gamma-2}}{\Gamma-1}
\end{equation}

and $a$ is either $a_1$ or $a_2$ depending on which region the EOS becomes acausal in.

\begin{figure*}[!t]
\centering
\resizebox{0.95\textwidth}{!}{\includegraphics{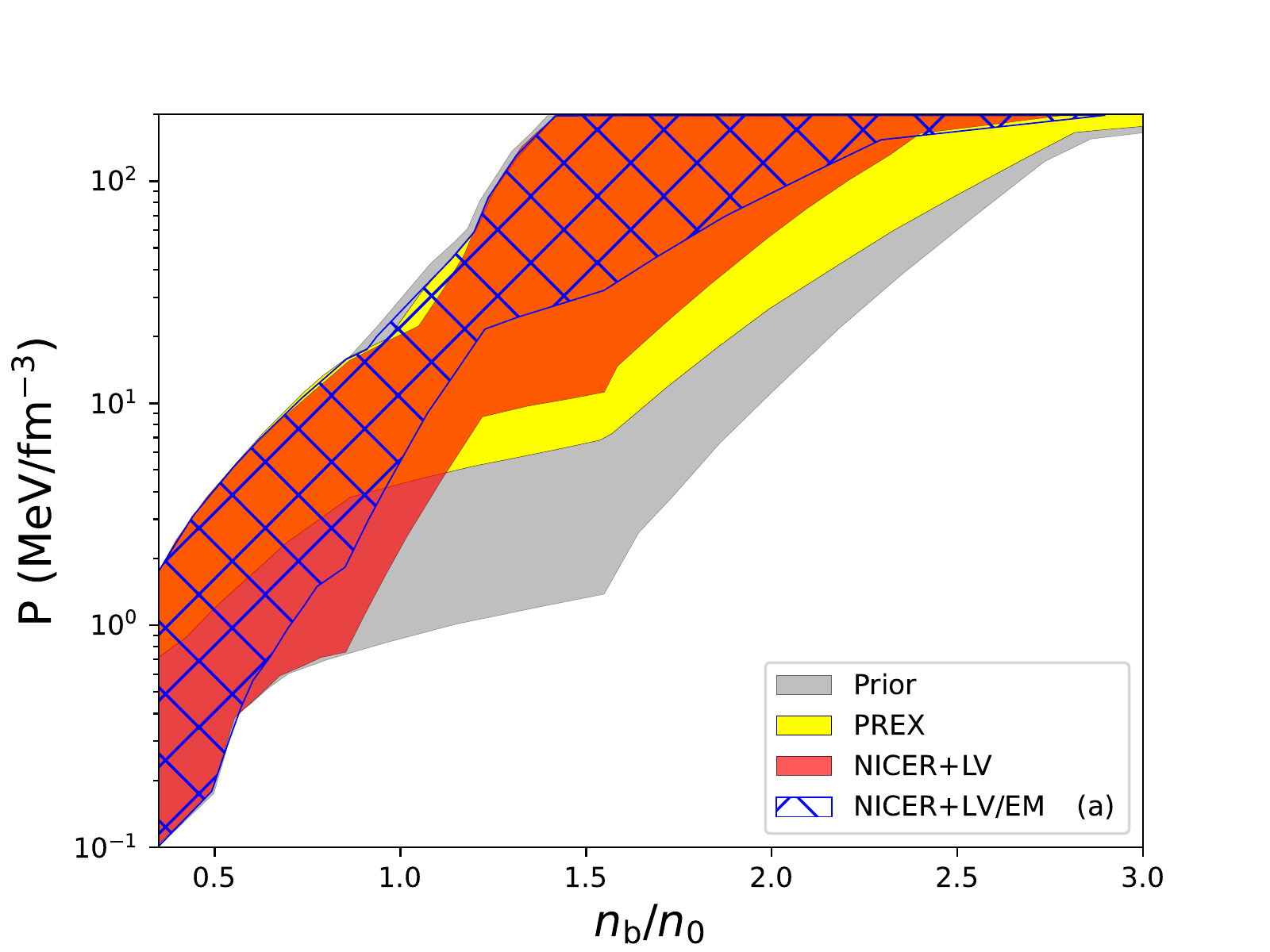}\includegraphics{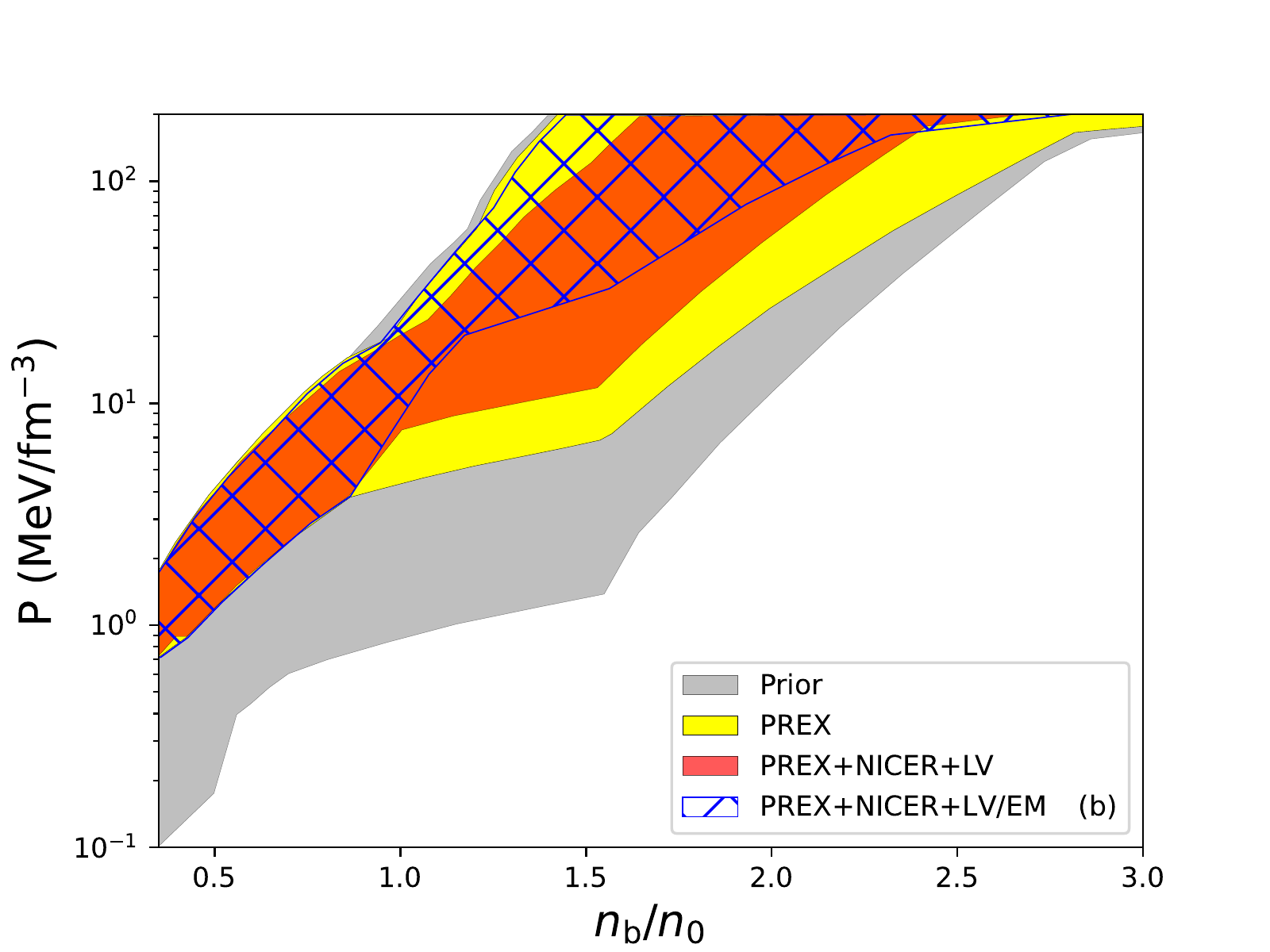}}
\resizebox{0.95\textwidth}{!}{\includegraphics{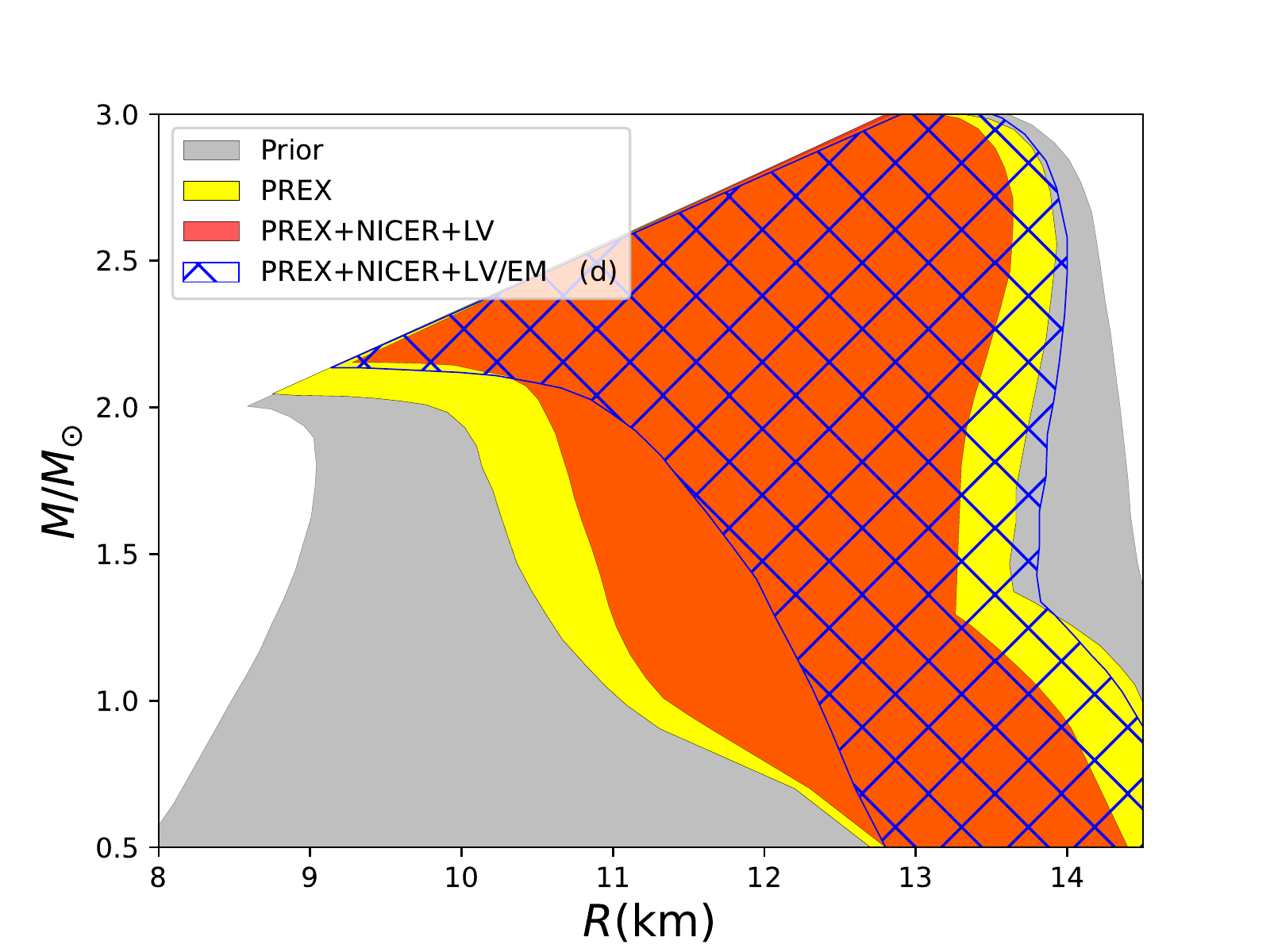}\includegraphics{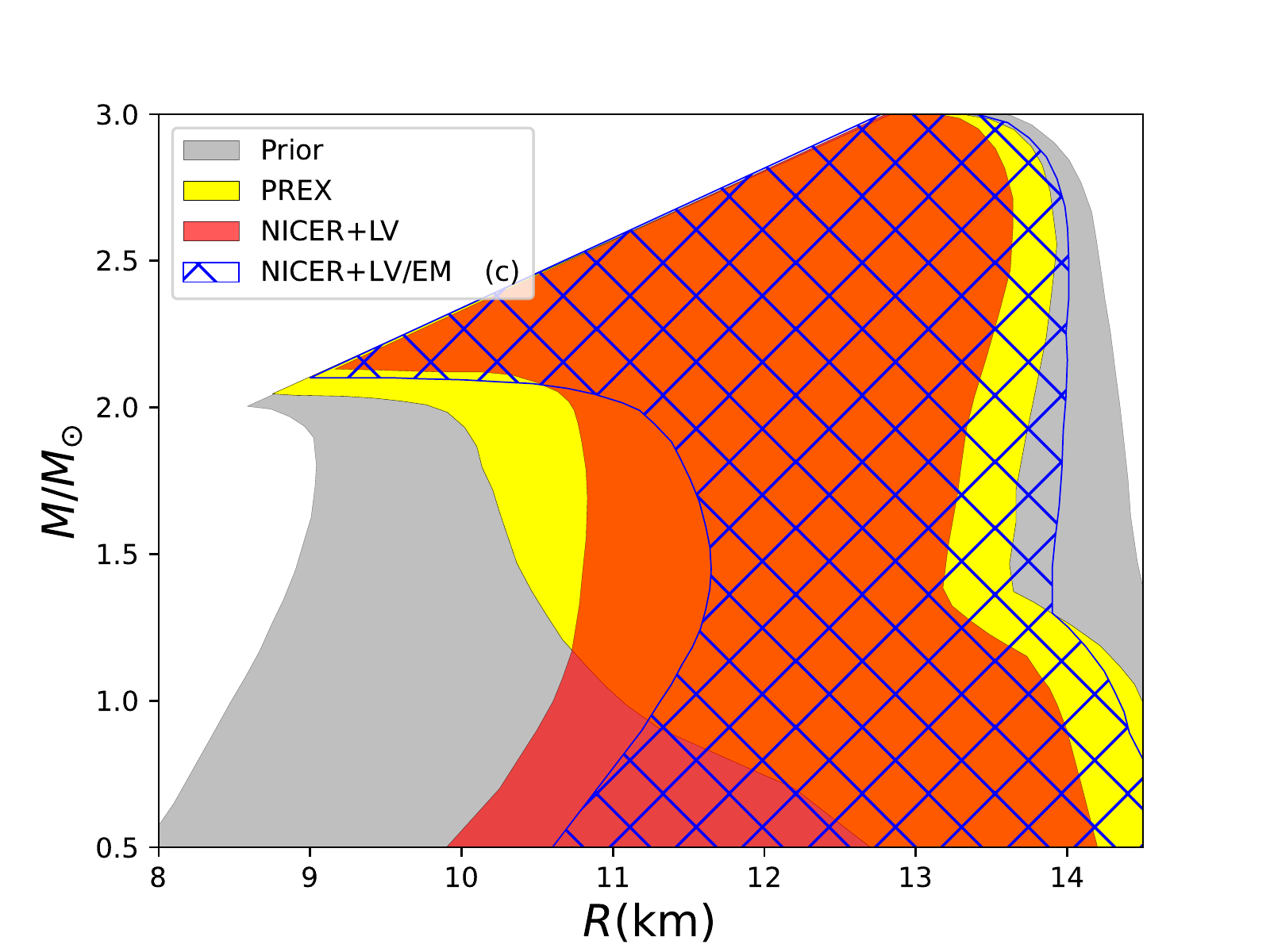}}
\caption{Equation of state ensembles use (a,b) and corresponding mass-radius curves (c,d) shown for our prior distribution (grey), with the inclusion of PREX data (yellow), NICER+LV data (red) and NICER+LV/EM data (blue hatched). The astrophysical data is combined with PREX in (b,d).}
\label{fig:1}       
\end{figure*}

Each equation of state generated is characterized by 5 parameters: the three symmetry energy coefficients $J$, $L$ and $K_{\rm sym}$ for the Skyrme-EOS, and the polytropic parameters $\Gamma_1$ and $\Gamma_2$. $\Gamma_2$, which controls the high density part of the EOS, can be tuned to give a desired maximum mass; $\Gamma_1$, which controls the EOS at intermediate densities in the core can be tuned to give a particular moment of inertia of a 1.4$M_{\odot}$ star $I_{1.338}$ while keeping the other parameters fixed. We can thus parameterize each EOS by $J$, $L$ and $K_{\rm sym}$, $I_{1.338}$, $M_{\rm max}$. The moment of inertia is not independent of tidal polarizability; a set of universal relationships between neutron star moments exists \cite{Yagi:2013aa,Yagi:2013ab,Yagi:2017aa}. Parameterizing the polytropes in terms of bulk properties of a neutron star allow us to incorporate other astrophysical information naturally. We know the maximum mass must be above 2 M$_{\odot}$ from measurements of the heaviest pulsars \cite{Demorest2010a-mass,Antoniadis2013a-mass,Raaijmakers:2021hc,Miller:2021hb}. Additionally, although we do not use this constraint here, modeling of GW170817 is suggestive of a maximum mass around 2.2 M$_{\odot}$ \cite{Shibata:2017aa,Margalit:2017aa,Ruiz:2018aa,Rezzolla:2018aa}.

We have already employed our ensemble to show how the multi-messenger detection of a gamma-ray flare coincident with the gravitational waves from the inspiral of a two neutron stars prior to merger allows a measurement of crust properties such as the average shear speed in the crust, from which symmetry energy constraints can be obtained \cite{Neill:2021lq}.

\section{Results}
\label{sec:results}

\begin{figure*}[!t]
\centering
\resizebox{0.95\textwidth}{!}{\includegraphics{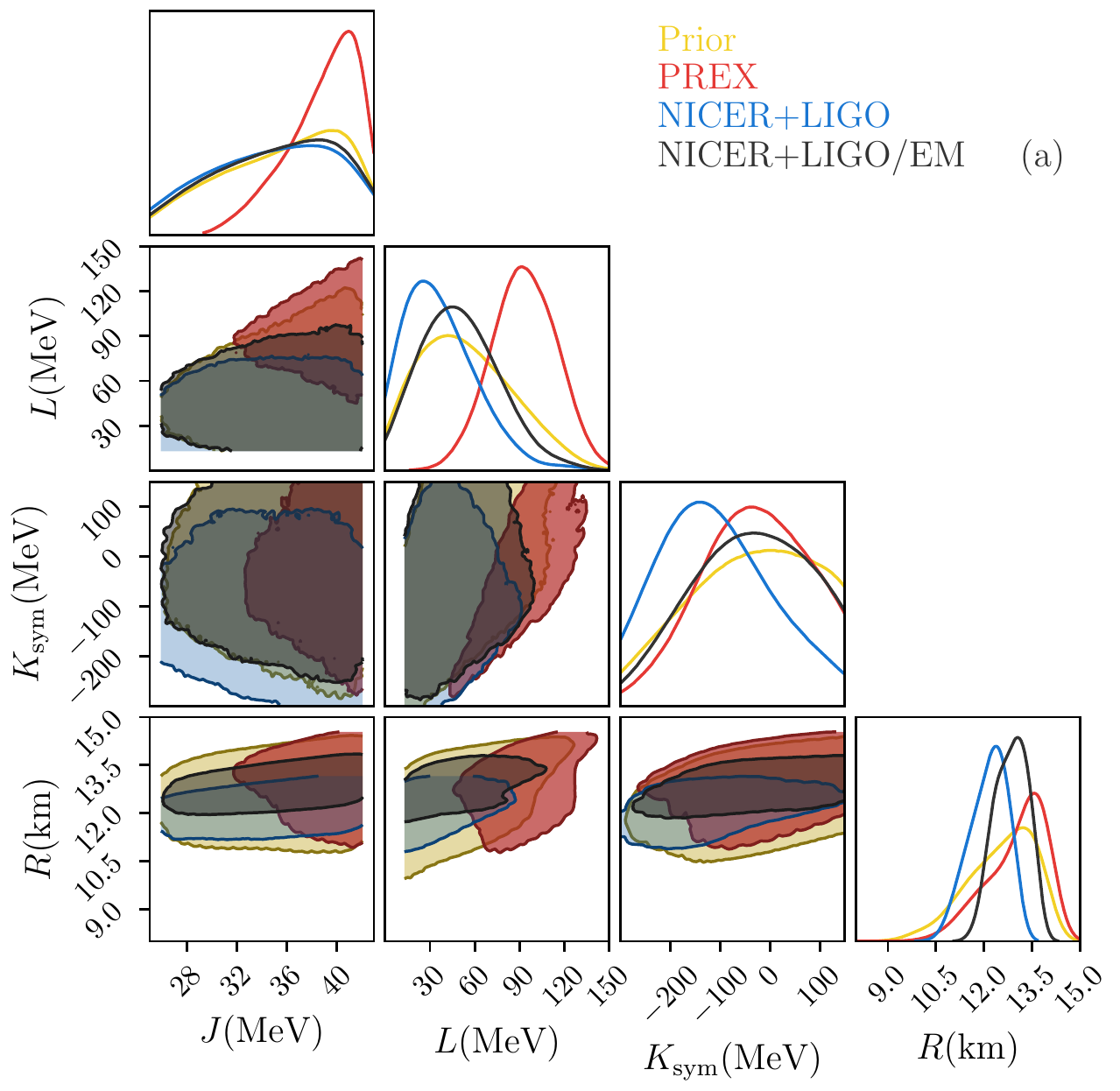}\includegraphics{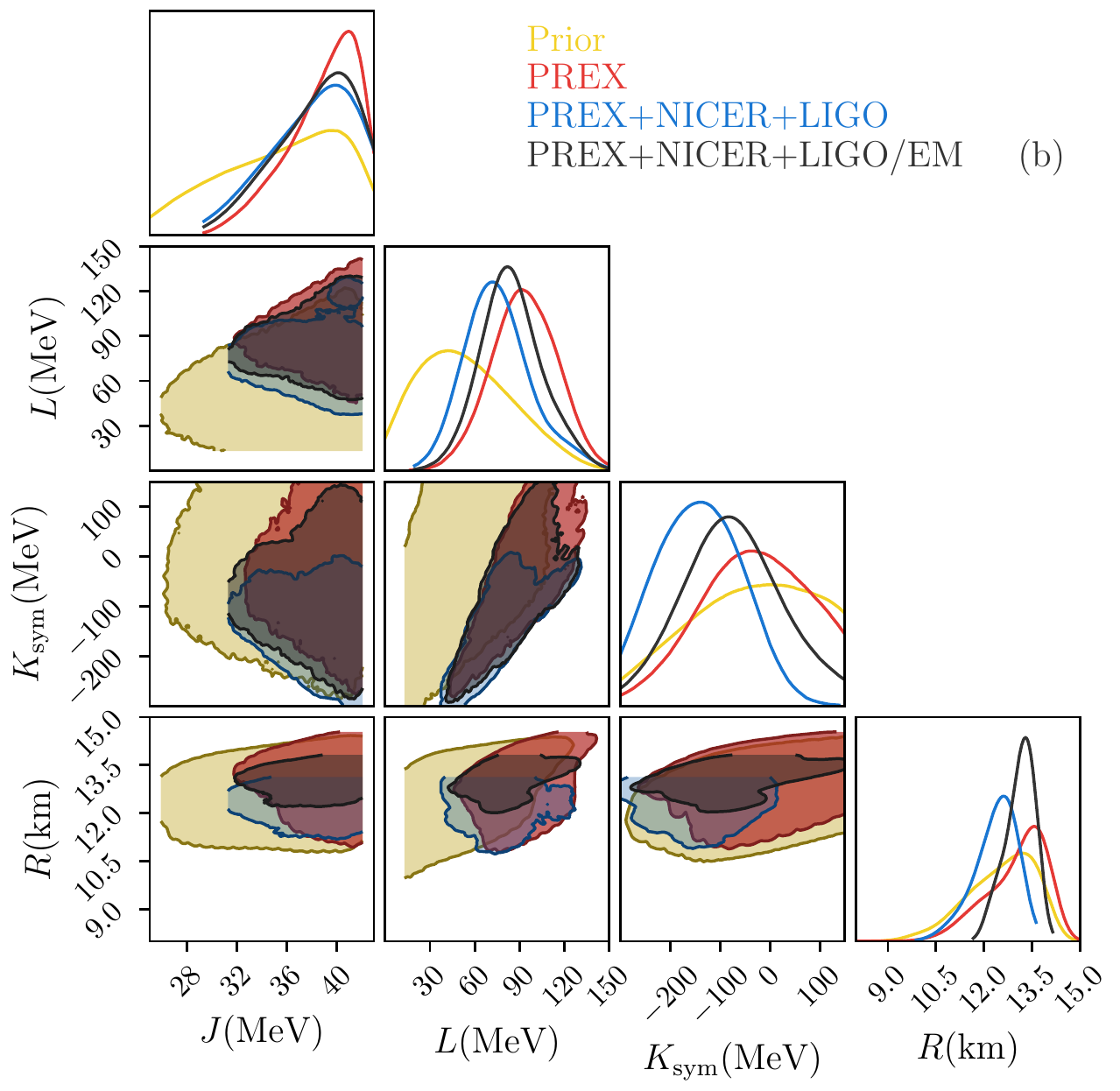}}
\caption{Marginalized distributions of our model predictions for the symmetry energy parameters $J$,$L$ and $K_{\rm sym}$, together with the radius of a 1.4M$_{\odot}$ star. Both plots give the results for the prior (Yellow) and PREX data (Red), with the NICER + LV and NICER + LV/EM data in blue and grey with (b) and without (a) the inclusion of PREX data.}
\label{fig:1}       
\end{figure*}

We construct a set of crust and outer core EOSs parameterized by $J$,$L$ and $K_{\rm sym}$ uniformly distributed on a 20$^3$ grid over the ranges $25<J<42$ MeV, $5<L<140$ MeV and $-450<K_{\rm sym}<150$ MeV (The ``uniform'' prior set from \cite{Balliet:2021lr}). These ranges are chosen to comfortably encompass all current experimental and theoretical constraints (which point to values of $J$ and $L$ in the range 26-34 MeV and 20-80 MeV respectively \cite{Lattimer:2014uq}), and also to allow for the possibility of a particularly stiff EOS hinted at by the PREX-II results which suggest a $J$ of 42 MeV and an $L$ of 140 MeV may be consistent with neutron skin data. Not all combinations of parameters produce viable crust models due to instabilities in the PNM EOS at low crust densities, so the symmetry energy space naturally gets filtered by this physical requirement. See \cite{Balliet:2021lr,Newton:2021tg} and Fig.~2 for the resulting parameter space for our priors. We then append two polytropes, with the parameters adjusted to give uniform priors on our maximum mass in the range 2.0$M_{\odot}<M_{\rm max}<3.0M_{\odot}$, since the NICER measurement of PSR J0740+6620 and radio observations of pulsars \cite{Demorest2010a-mass,Antoniadis2013a-mass,Raaijmakers:2021hc,Miller:2021hb} show neutron stars above two solar masses are possible. The upper bound of 3.0$M_{\odot}$ is below the limit set by causality, and while we could find higher masses allowed by causality in certain regions of parameter space, we choose for this exploratory work to set this limit. Our priors on the moment of inertia of a 1.338$M_{\odot}$ star are uniform between the smallest and largest physically possible for a given $J,L,K_{\rm sym}$ and $M_{\rm max}$. This results in an ensemble of about 60,000 EOSs. 

Although the ensembles of EOSs are designed for a full Bayesian inference combining nuclear and astrophysical data, the results of which are in preparation, the purpose of this paper is to demonstrate in a simple way the fruitfulness of the approach. We do this by applying some nuclear and astrophysical constraints as a filter on a prior ensemble of EOSs. We apply the following filters to the EOSs. For each filter, we simply remove the EOSs that do not fulfill the constraint. Although it will sometimes be convenient to present the results as a median and credible interval, keep in mind these do not come from a full statistical inference.

The filters are: 

\begin{enumerate}
    \item The PREX 68\% confidence limits $0.21<r_{\rm np}^{208}<0.35$ fm \cite{Adhikari:2021tm}.
    \item The LIGO 68\% confidence limits on the mass-weighted tidal deformability $70 < \Tilde{\Lambda} < 580$ \cite{Abbott:2018fe}.
    \item Since the above constraint tends to favor softer EOSs, we also include the constraints from combining the GW data with EM information about the amount of ejecta from GW170817 \cite{Radice:2019xy}, resulting in a lower bound on the tidal deformability $300 < \Tilde{\Lambda} < 800$. This gives us an example of an astrophysical dataset that favors stiffer EOSs, to compare to the PREX dataset. 
    \item The 68\% confidence regions of the masses and radii of two X-ray pulsars from NICER X-ray timing. We filter our EOSs through both the combined the 68\% confidence limits for the radius of pulsar J0030+0451 \cite{Riley:2019aa,Miller:2019jt} and PSR J0740+6620 \cite{Miller:2021hb,Riley:2021sp}.
\end{enumerate}

\noindent We label the PREX filter as \emph{PREX}, the astrophysical filter combining NICER and LIGO/VIRGO data (datasets 2 and 4 above) as \emph{NICER+ LV}, the astrophysical datasets including the extra information from EM observations of the kilonova (datasets 3 and 4 above) as \emph{NICER+ LV/EM}, and the combination of the astrophysical and neutron skin datasets as \emph{PREX+ NICER+ LV} and \emph{PREX+ NICER+ LV/EM}.

The 68\% confidence limits from LIGO and VIRGO are obtained from a Bayesian analysis which marginalizes over a number of other variables - for example, distances to the systems, orbital inclination in the case of LIGO, the  parameters characterizing the geometry of the hot spot on the neutron stars observed by NICER. Despite this, we should keep in mind that there may be additional uncertainty - particularly systematic - that has yet to be characterized. For example, the constraint on the neutron skin of lead from PREX was obtained by an analysis using a small number of relativistic mean field models. 

In Fig.~1, we show the EOS space covered by our priors and for the filtered sets of EOSs. In Fig.~1a we compare the priors (grey) with the PREX filter alone (yellow), the NICER+LV filters without PREX (red), and the NICER+ LV/EM filters without PREX (blue hatched). In Fig.~1b we show again the Prior and PREX filters for reference, this time with the NICER+LV and NICER+ LV/EM filters combined with the PREX filter. The resulting mass-radius relations are shown in Figs.~2a and 2b for the same combinations of filters. In both figures, the shaded areas bound all our EOSs for a given filter, and their mass-radius curves.

The PREX data eliminates the softest EOSs, and has its most pronounced effect on low mass neutron stars where the EOS at nuclear saturation density is most influential. Indeed, it eliminates the softest \emph{crust} EOSs, below 0.5$n_0$. The priors give radii ranging from just below 9km up to just over 14km, with the PREX data predicting lower limits of radii between 10 and 11km above one solar mass, increasing to above 12km for neutron stars significantly below 1 M$_{\odot}$.

The NICER+LV data allows for soft EOSs close to $n_0$, but then eliminates EOSs that are soft at high densities so that stars above 2$M_{\odot}$ are common in our ensemble. It therefore provides complementary information to PREX. Finally, the NICER+ LV/EM data further eliminates soft EOSs, particularly at high densities, due to the lower bound on the tidal deformability. One can see that this stiff-EOS favoring astrophysical dataset retains many soft EOSs at lower density, but eliminates many of them around saturation density (though not as many as PREX) and above, where the PREX data allows softer EOSs. This again reveals the complementarity of PREX and GW data; we are in a stronger position when we can combine both consistently.

\begin{figure*}[!t]
\centering
\resizebox{0.78\textwidth}{!}{\includegraphics{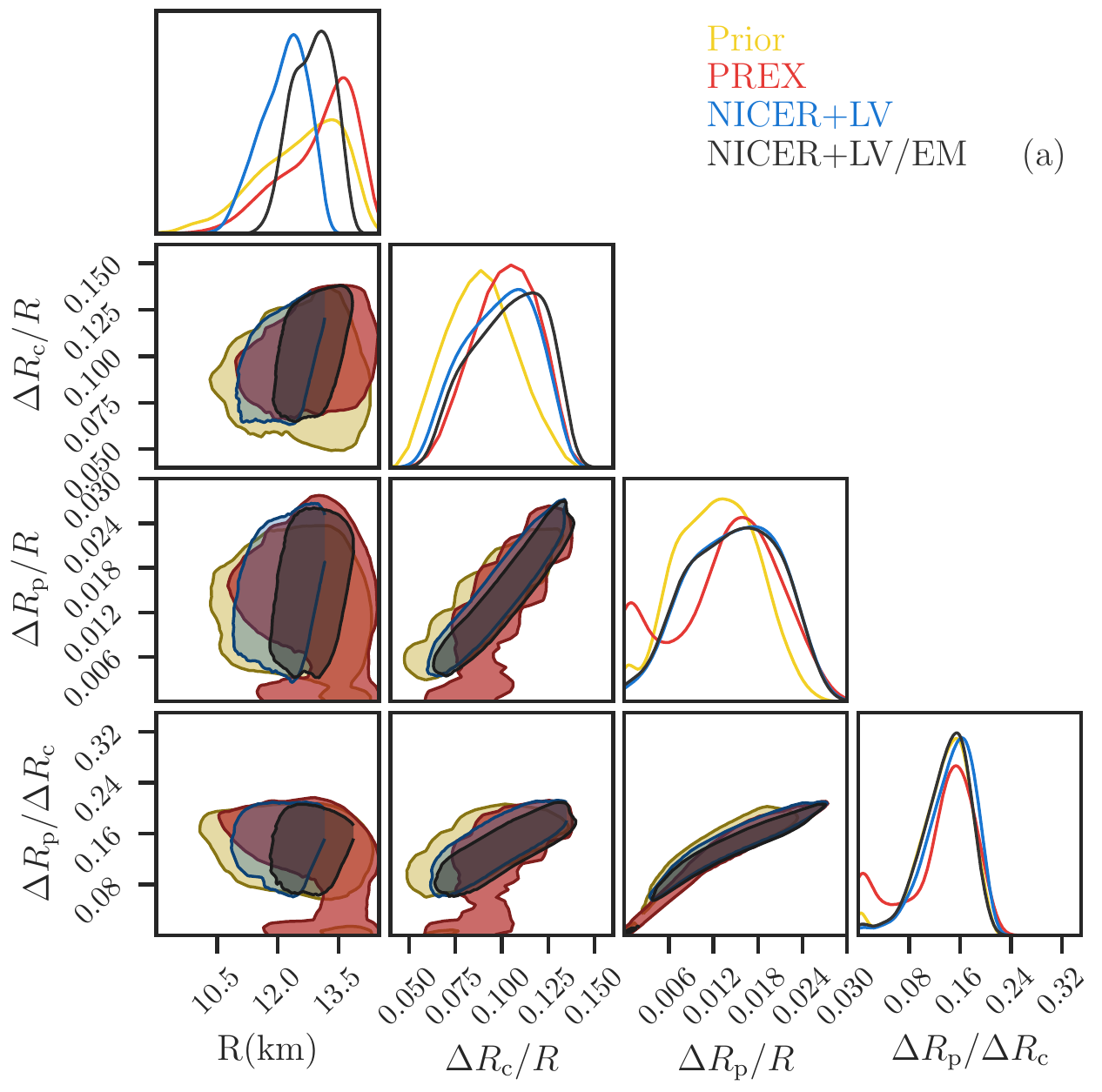}\includegraphics{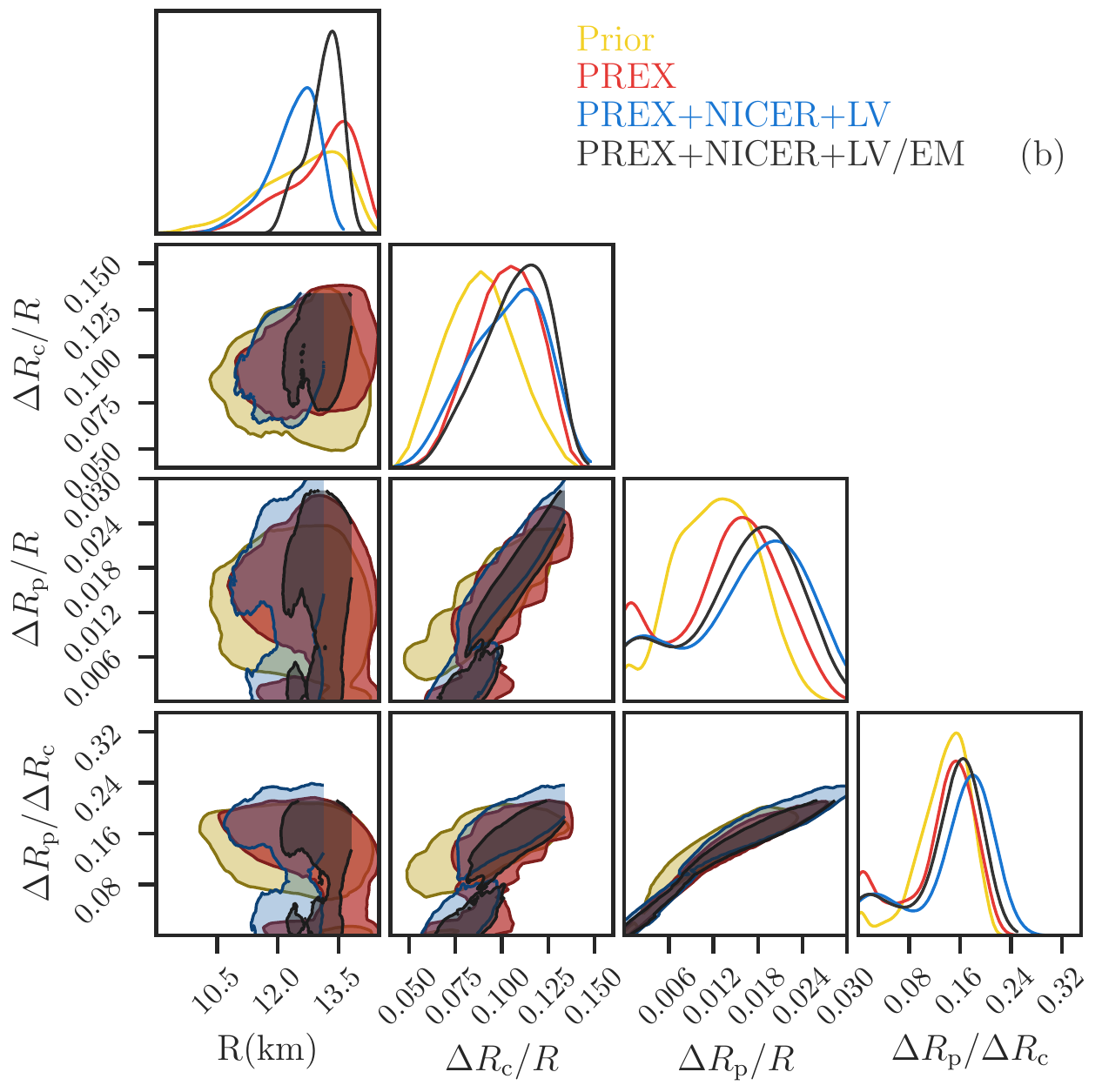}}
\resizebox{0.78\textwidth}{!}{\includegraphics{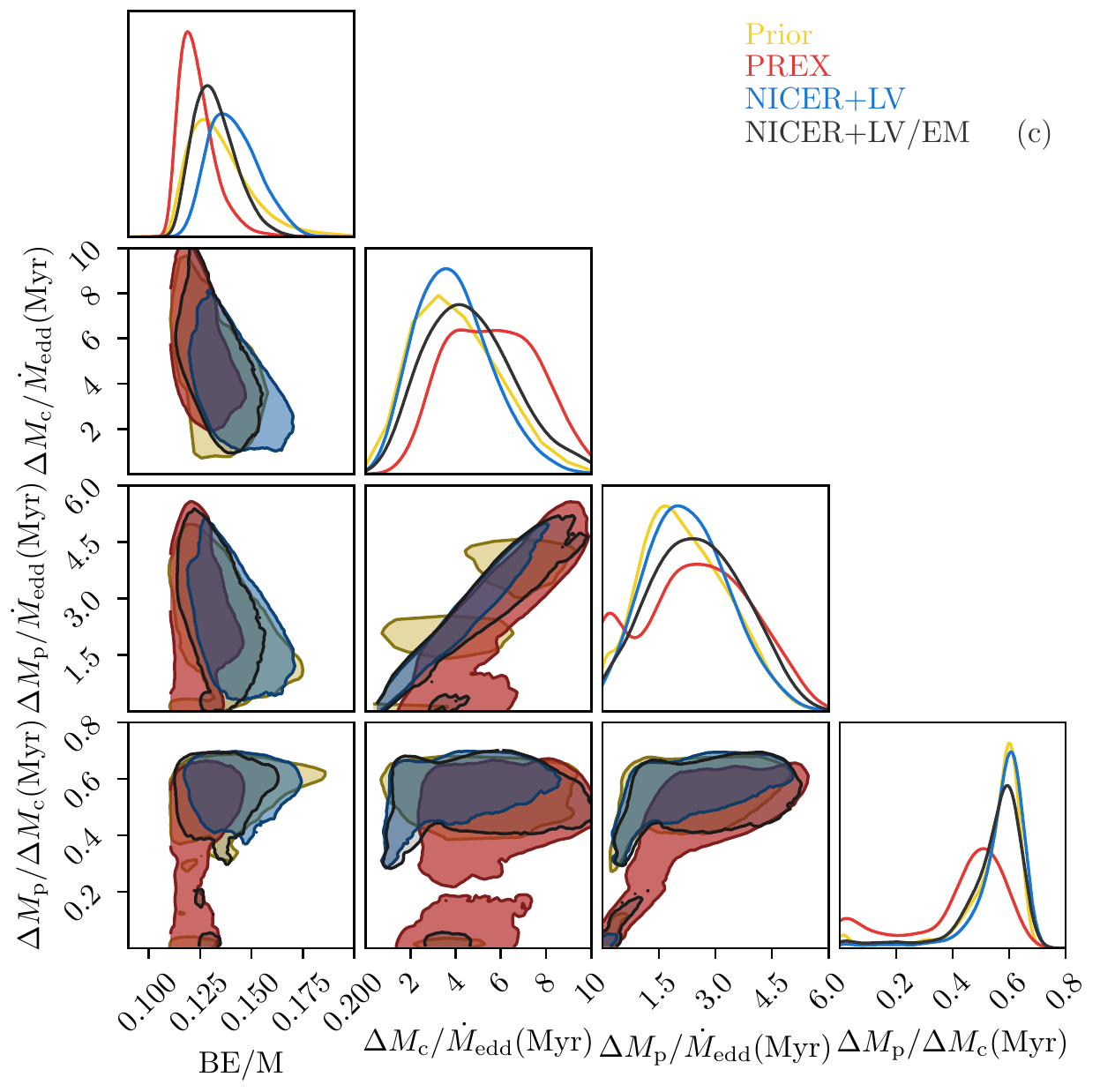}\includegraphics{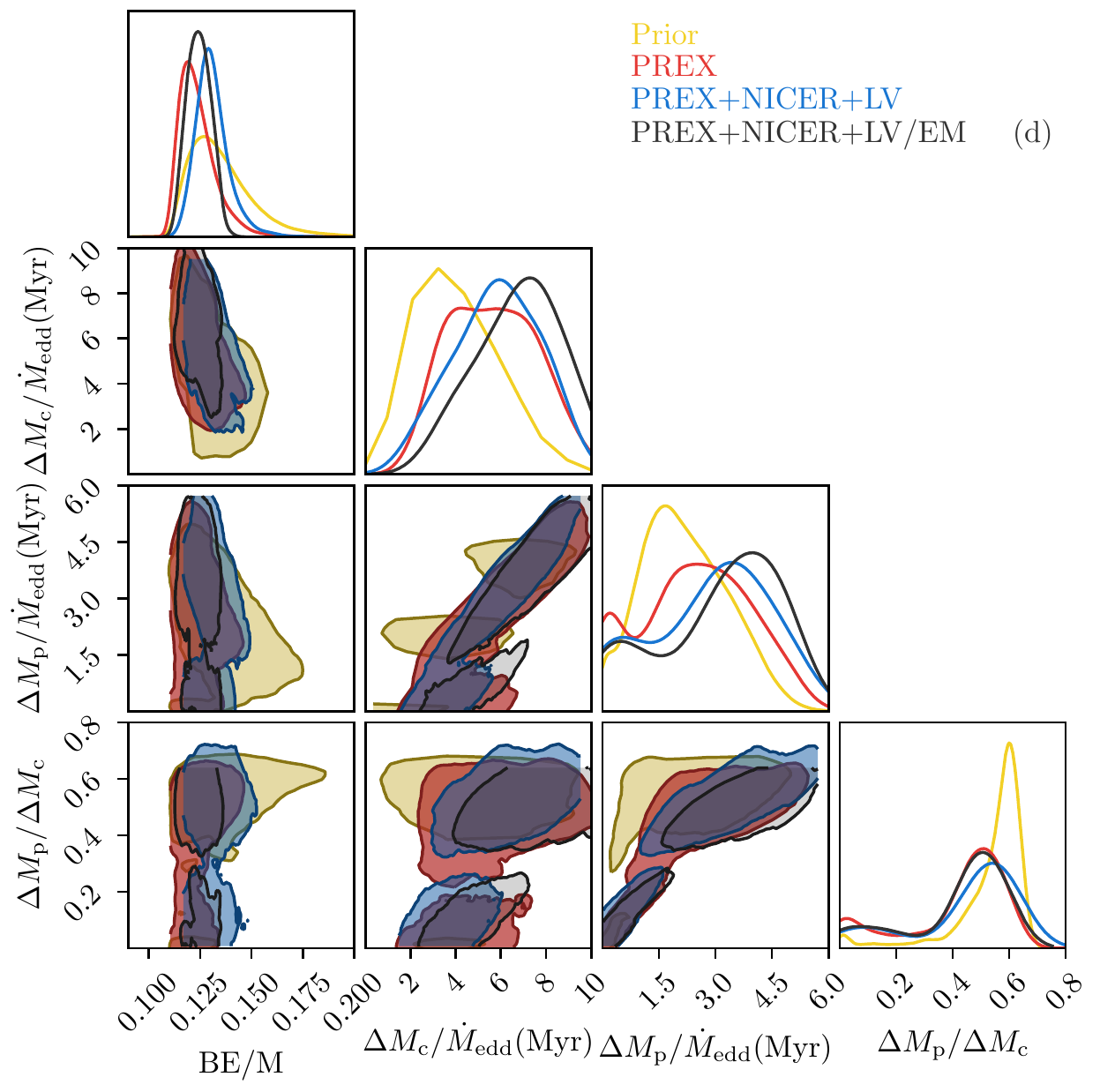}}
\resizebox{0.78\textwidth}{!}{\includegraphics{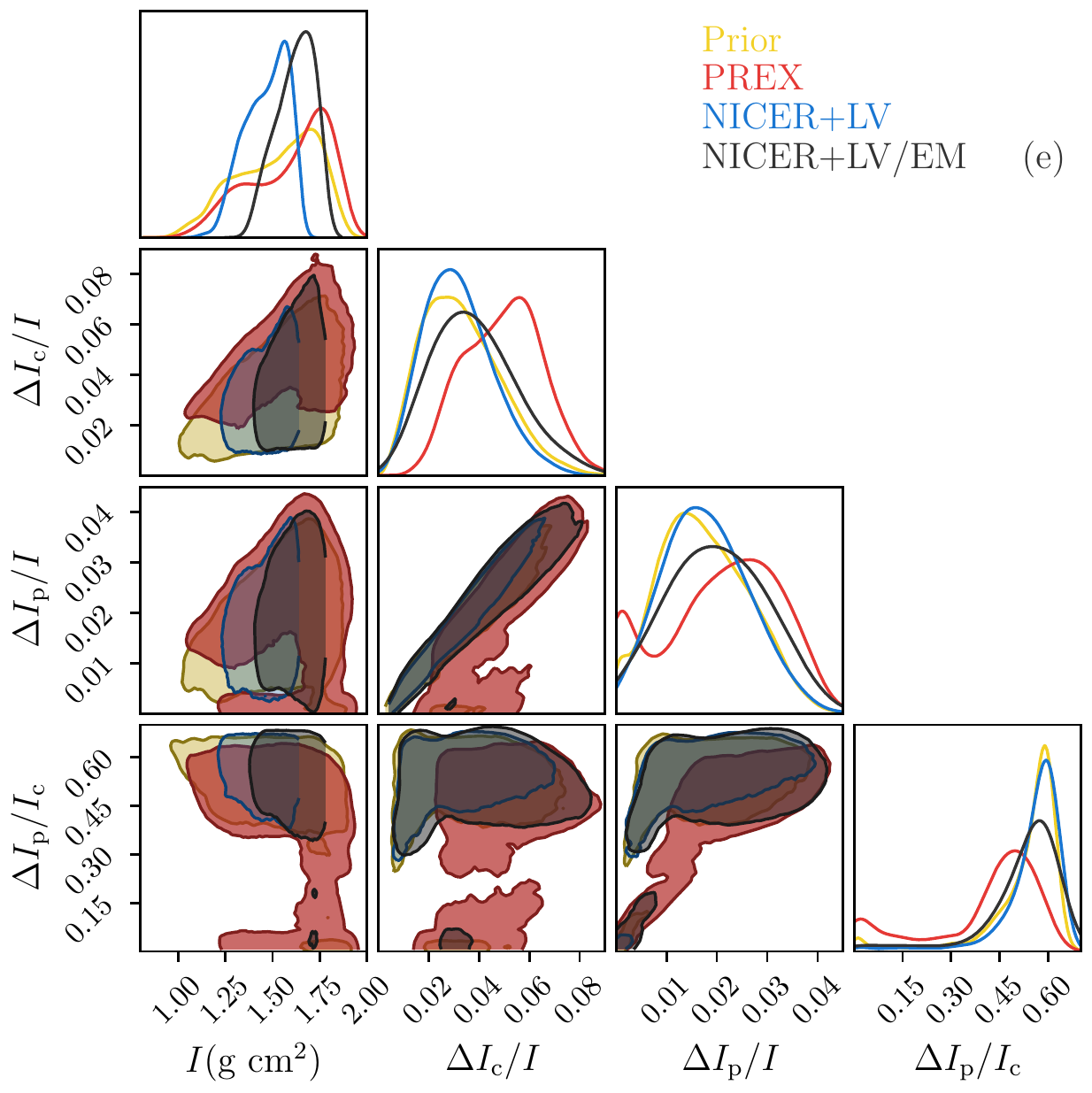}\includegraphics{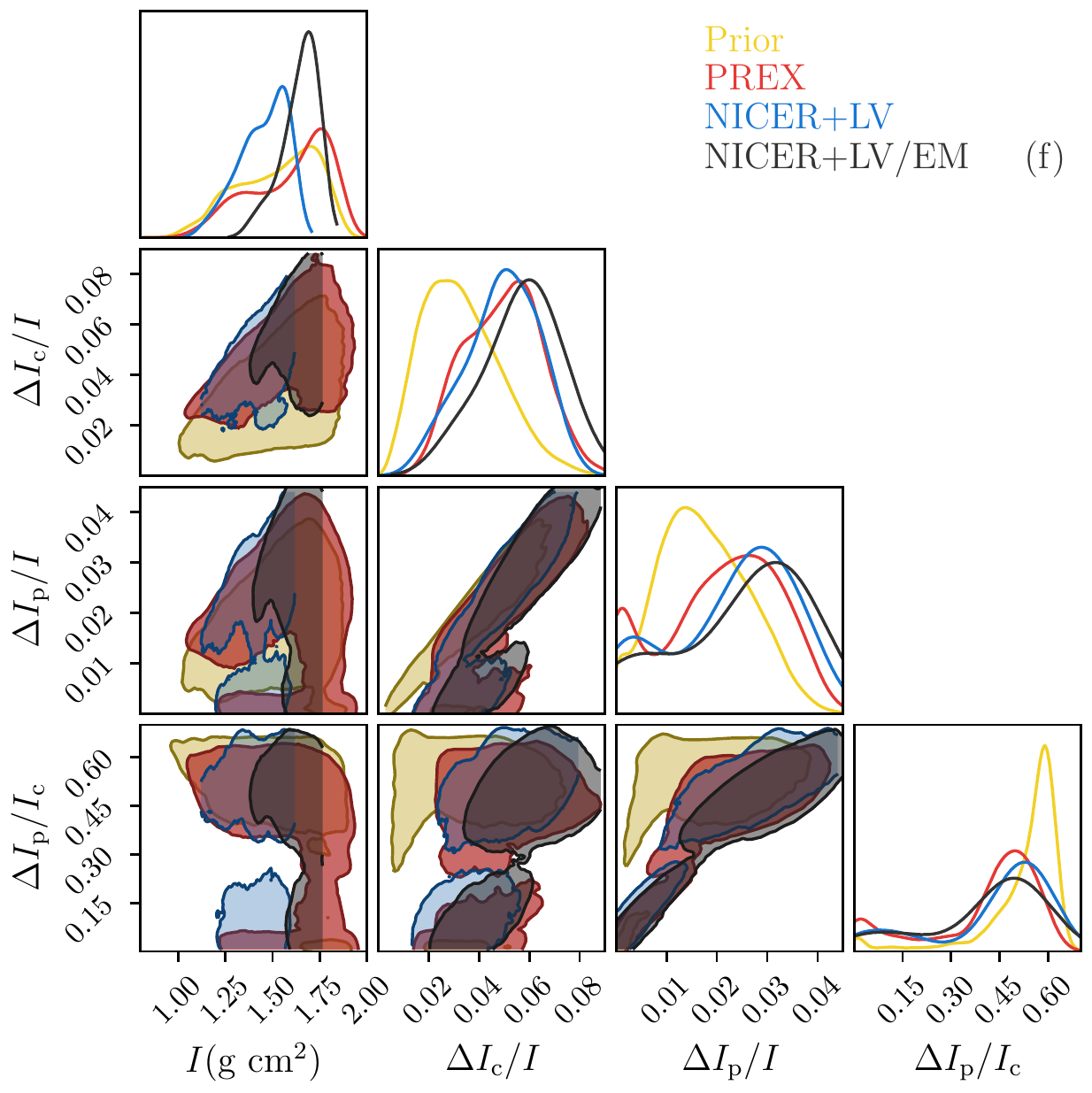}}
\caption{Marginalized distributions of our model predictions for radii and crust and pasta thicknesses (a,b), binding energy and crust and pasta masses (c,d) and moment of inertia of a 1.338 M$_{\odot}$ data and of the crust and pasta (e,f). The crust and pasta masses are given as a crust replacement timescale in Myr by dividing by the Eddington accretion rate. All quantities are for a 1.4 M$_{\odot}$ star unless otherwise specified. All plots give the results for the prior (Yellow) and PREX data (Red), with the NICER+LV and NICER+ LV/EM data in blue and grey with (b,d,f) and without (a,c,e) the inclusion of PREX data.}
\label{fig:1}       
\end{figure*}

When NICER+LV data is combined with PREX data (Fig.~1b and 2b), the complementary nature of the nuclear and astrophysical constraints are clearly seen, with soft EOSs at both low densities and high densities being eliminated by different components of the dataset, and correspondingly, larger radii predicted for lower and higher mass neutron stars.

In Fig.~2a and 2b, we plot regions in symmetry energy space where 90\% of our models lie after filters are applied. In Table~1 the numerical predictions for all quantities discussed in this paper are given, as medians and 68\% limits about the median after the filters are applied. $J$ is constrained most by the PREX data, to large values $39.4^{+2.7}_{-3.6}$ MeV. Compared to the prior value of $53^{+32}_{-32}$ MeV, $L$ is constrained by PREX to much stiffer values $93^{+24}_{-24}$ MeV and The NICER+LV data alone predicts a much softer value $29^{+32}_{-16}$ MeV. Additionally, the NICER+LV data has the largest effect on $K_{\rm sym}$ shifting the prior prediction of $-2^{+111}_{-148}$ MeV to lower values of $-113^{+111}_{-111}$ MeV. The extra EM data has little effect on the symmetry energy inference, emphasizing that its stiffening effects occurs at high densities.

Note that these confidence limits are obtained marginalizing over all other parameters. The 68\% ranges about the median obtained from certain data may be outside the 68\% ranges for the priors, as the data doesn't constrain each parameter separately, rather the five dimensional parameter space where they live.

When the PREX data is combined with the astrophysical data, there is little difference in the resulting medians and ranges of $J$, $L$, with the PREX having the dominant effect; there is a difference between the predicted ranges of $K_{\rm sym}$, with NICER+ LV+ PREX data shifting the median from -113 MeV to -150 MeV compared to the median from PREX of -39 MeV. Adding PREX to the NICER+LV data also shrinks the 68\% range of $K_{\rm sym}$ from 220 MeV down to around 150 MeV. NICER+LV data is more effective at constraining $K_{\rm sym}$ than the PREX data at the current level of precision.

\begin{table*}[!th]
\caption{Medians and 68\% ranges about the median for all quantities plotted in Figs.~2 and ~3, for each data set combination.}
\label{tab:1}       
\begin{tabular}{p{26mm}p{18mm}p{18mm}p{18mm}p{24mm}p{24mm}p{24mm}}
\hline\noalign{\smallskip}
Quantity & Prior & PREX & NICER+\newline LIGO & NICER\newline+LV/EM & PREX+NICER\newline+LV & PREX+NICER \newline+LV/EM \\
\noalign{\smallskip}\hline\noalign{\smallskip}
$J$(MeV) & $ 36.7^{+4.5}_{-6.3}$ & $ 39.4^{+2.7}_{-3.6}$ & $ 34.9^{+5.4}_{-5.4}$ & $ 35.8^{+4.5}_{-6.3}$ & $39.4^{+2.6}_{-4.5}$ & $39.4^{+1.8}_{-4.5}$\\
$L$(MeV) & $ 53^{+32}_{-32}$ & $ 93^{+24}_{-24}$ & $ 29^{+32}_{-16}$ & $ 45^{+32}_{-24}$ & $ 79^{+16}_{-16}$ & $ 85^{+16}_{-16}$ \\
$K_{\rm sym}$ (MeV) &  $ -2^{+111}_{-148}$ & $ -39^{+148}_{-111}$ & $ -113^{+111}_{-111}$ & $ -39^{+148}_{-111}$ & $ -150^{+74}_{-74}$ &  $ -76^{+74}_{-74}$\\
$R$ (km) &  $ 12.7^{+0.9}_{-1.3}$ & $ 13.3^{+0.6}_{-1.2}$ & $ 12.2^{+0.5}_{-0.7}$ & $ 12.9^{+0.5}_{-0.6}$ & $ 12.5^{+0.5}_{-0.7}$ &  $ 13.2^{+0.3}_{-0.6}$\\
$\Delta R_{\rm c}/R$ & $ 0.088^{+0.019}_{-0.019}$  & $ 0.103^{+0.017}_{-0.018}$ & $ 0.100^{+0.018}_{-0.022}$ & $ 0.105^{+0.019}_{-0.023}$ & $ 0.105^{+0.017}_{-0.023}$ &  $ 0.109^{+0.016}_{-0.020}$\\
$\Delta R_{\rm p}/R$ & $ 0.013^{+0.005}_{-0.006}$  & $ 0.015^{+0.006}_{-0.011}$ & $ 0.015^{+0.006}_{-0.007}$ & $ 0.015^{+0.006}_{-0.007}$ & $ 0.018^{+0.006}_{-0.007}$ &  $ 0.017^{+0.006}_{-0.012}$\\
$\Delta R_{\rm c}/\Delta R_{\rm p}$ &  $0.141^{+0.032}_{-0.044}$ & $ 0.141^{+0.036}_{-0.101}$ & $ 0.150^{+0.030}_{-0.043}$ & $ 0.141^{+0.030}_{-0.042}$ & $ 0.172^{+0.025}_{-0.124}$ &  $ 0.155^{+0.027}_{-0.097}$\\
$BE/M$ & $ 0.133^{+0.018}_{-0.012}$ & $ 0.122^{+0.010}_{-0.006}$ & $ 0.140^{+0.014}_{-0.011}$ & $ 0.131^{+0.011}_{-0.009}$ & $ 0.130^{+0.008}_{-0.005}$ &  $0.124^{+0.006}_{-0.006}$ \\
$\Delta M_{\rm c}$/$\dot{M}_{\rm edd}$ (Myr) &  $ 3.7^{+2.2}_{-1.6}$ & $ 5.6^{+2.1}_{-2.1}$ & $ 3.8^{+1.8}_{-1.5}$ & $ 4.5^{+2.1}_{-1.8}$ & $ 5.9^{+2.0}_{-2.0}$ &  $ 7.0^{+1.9}_{-2.2}$\\
$\Delta M_{\rm p}$/$\dot{M}_{\rm edd}$ (Myr) & $ 2.0^{+1.3}_{-1.0}$ & $ 2.5^{+1.5}_{-1.9}$ & $ 2.2^{+1.2}_{-1.0}$ & $ 2.5^{+1.3}_{-1.3}$ & $ 3.1^{+1.3}_{-2.6}$ &  $ 3.4^{+1.2}_{-2.6}$\\
$\Delta M_{\rm c}/\Delta M_{\rm p}$ &  $ 0.58^{+0.05}_{-0.12}$ & $ 0.47^{+0.09}_{-0.34}$ & $ 0.59^{+0.04}_{-0.09}$ & $ 0.57^{+0.05}_{-0.11}$ & $ 0.51^{+0.08}_{-0.36}$ &  $ 0.48^{+0.09}_{-0.31}$\\
$I$(10$^{45}$g cm)$^2$ & $ 1.56^{+0.19}_{-0.28}$ & $ 1.65^{+0.15}_{-0.31}$ & $ 1.48^{+0.10}_{-0.14}$ & $ 1.63^{+0.09}_{-0.12}$ & $ 1.47^{+0.10}_{-0.16}$ &  $ 1.67^{+0.06}_{-0.011}$\\
$\Delta I_{\rm c}/I$ & $ 0.032^{+0.018}_{-0.014}$ & $ 0.050^{+0.014}_{-0.017}$ & $ 0.031^{+0.015}_{-0.012}$ & $ 0.037^{+0.018}_{-0.015}$ & $ 0.050^{+0.014}_{-0.016}$ &  $ 0.058^{+0.013}_{-0.017}$\\
$\Delta I_{\rm p}/I$ & $ 0.017^{+0.010}_{-0.008}$ & $ 0.022^{+0.010}_{-0.016}$ & $ 0.017^{+0.009}_{-0.008}$ & $ 0.020^{+0.010}_{-0.010}$ & $ 0.026^{+0.009}_{-0.021}$ &  $ 0.028^{+0.090}_{-0.022}$\\
$\Delta I_{\rm c}/\Delta I_{\rm p}$ & $ 0.57^{+0.04}_{-0.12}$ & $ 0.46^{+0.09}_{-0.33}$ & $ 0.58^{+0.04}_{-0.09}$ & $ 0.56^{+0.04}_{-0.11}$ & $0.50^{+0.08}_{-0.35}$ &  $ 0.46^{+0.08}_{-0.32}$ \\
\noalign{\smallskip}\hline
\end{tabular}
\end{table*}

In Figs.~3 we show the 90\% regions for our model predictions of the nuclear pasta layer (3a,b), their mass (3c,d) and their moment of inertia (3e,f) compared with the radius, binding energy and moment of inertia of the star respectively, all for a canonical 1.4M$_{\odot}$ star except the moment of inertia which is given for a 1.338 M$_{\odot}$ star. In Figs.3(a,c,e) we show the results from the priors and PREX filters compared to the astrophysical data NICER+LV and NICER+LV/EM. In Figs.3(b,d,f) we show the results from the priors and PREX filters compared with the combined PREX and astrophysical data. The crust and pasta masses are expressed in astrophysical relevant units in terms of the Eddington mass accretion rate $\dot{M}_{\rm edd}$; they are thus expressed as crust replacement timescale assuming accretion at Eddington rates.

For the bulk properties, for $R$ and $I$, the constraining power of the astrophysical data is much greater than that of the PREX data, and vice-versa for binding energy $BE/M$. When PREX and NICER+LV data are applied, the distributions for $BE/M$ shift to closely match the PREX distribution alone. For $R$ and $I$, adding the PREX data does not significantly affect the distributions. 
For $R$ and $I$, the astrophysical data (expectedly) narrows the uncertainty to a much greater degree than the PREX data. The NICER+LV/EM and PREX data give similar median values for $R$ and $I$, with the astrophysical data narrowing the uncertainty significantly and the nuclear data having little effect on the uncertainty. PREX predicts relatively large radii $13.3^{+0.6}_{-1.2}$km, and adding the NICER+LV data brings the prediction back down to $12.5^{+0.5}_{-0.7}$ km data alone. A similar behavior is seen for the moment of inertia.

The PREX and NICER+LV data lower and raise the median binding energy by comparable amounts $\approx 8\%$. The PREX data decreases the 68\% width by about half, while the NICER+LV data shrinks it by about 25\%.

Moving on to the crustal properties, note the astrophysical data and PREX data give similar information on the crust and pasta thickness. Adding the PREX, NICER+ LV and EM data sequentially systematically increases the median crust and pasta thicknesses by 20-25\%. The most significant data effect on the relative thickness of pasta and crust layers is for the combination of the PREX+NICER+ LV data, which increases the median relative pasta thickness by around 20\% compared to the prior; the other data have a much smaller effect.

Compared to the priors, the PREX data increases the median crust mass and moment of inertia by about 50\%, with a smaller 20\% increase for the pasta mass and moment of inertia. PREX data decreases the median pasta fraction of the crust by almost 20\% compared to the prior; the astrophysical datasets alone do not affect the median values significantly. However, combining all the datasets we get the biggest effect, with the medians shifting by over 50\% to large masses and moments of inertia of the crust.

In \cite{Balliet:2021lr,Newton:2021tg} we used an approximation to find the relative mass of pasta in the crust.

\begin{equation}
\frac{\Delta M_{\rm p}}{\Delta M_{\rm c}} \approx \frac{P_{\rm p}}{P_{\rm cc}} \;\;\;\;\;\;\; \frac{\Delta R_{\rm p}}{\Delta R_{\rm c}} \approx
\frac{\mu_{\rm cc} - \mu_{\rm p}}{\mu_{\rm cc}-\mu_0}.
\end{equation}

\noindent where $P_{\rm c,p}$, $\mu_{\rm c,p}$ are the crust-core and pasta transition pressures and chemical potentials respectively, and $\mu_0$ is the chemical potential at the surface of the star. In a future work we will conduct a more thorough comparison of the exact values obtained here with these approximate relations. Here we note that, taking into account the difference in analysis (full Bayesian versus a simple filter here), the results for the thickness and mass of the pasta layers relative to the crust are consistent with the approximation.

Neither astrophysical data nor PREX data narrow the uncertainty in the crust and pasta thicknesses, mass or moments of inertia; indeed, the uncertainty appears to increase significantly with the PREX data, an artifact of the increased likelihood of no pasta causing a second peak in the distribution \cite{Newton:2021tg}. Thus currently data is giving information about the most likely value, but not increasing the precision of the prediction.

\section{Discussion}
\label{Discussion}

Let us discuss our results in the context of similar studies that have attempted to combine PREX data with astrophysical measurements, or build unified crust-core EOS ensembles.

A number of studies have incorporated neutron skin data using the universal relation with $L$ \cite{Roca-Maza:2011aa}. In \cite{Essick:2021vn}, Gaussian processes (GPs) are used to construct the EOS, trained on EDFs but with large uncertainties on the GP hyperparameters so that a very wide range of EOS space is explored. Although we do not explore a comparable parameter space at high densities - a simple two-polytrope model does not find multiple stable branches, for example - we do end up covering a similar region of parameter space in ($J,L,K_{\rm sym}$) space (compare Fig.~2 here with Fig.~3 from \cite{Essick:2021vn}). We obtain similar correlations when comparing $L$ to $R$; our median values for the symmetry energy parameters are somewhat different for the astrophysical data and the astrophysical data combined with PREX, apart from the prediction of $L$ for which both studies obtain a median of 80 MeV. These differences could be due to our simplified filtering of models by data, the simplified treatment of the neutron skin thickness of lead in \cite{Essick:2021vn}, or the different ranges for the prior distribution of symmetry energy parameters; this will be examined in future work.

The nuclear matter EOS + 3 piecewise polytrope method used by \cite{Biswas:2021pd}, obtain 12.21$<R<13.17$km to 1$\sigma$ confidence (compare with our result 12.1$<R<$13.9km from PREX data alone), together with symmetry energy ranges of 49.53$<L<$89.47 MeV and -330.62$<K_{\rm sym}<$-0.57 MeV. These compare with our ranges of 69$<L<$117MeV and -224$<K_{\rm sym}<$-76 MeV. Using a similar method, \cite{Tang:2021rt} obtain 52$<L<$91 MeV and -260$<K_{\rm sym}<$13 MeV.

Constraints on $R_{1.4}$ from NICER and PREX measurements were obtained for a limited set of RMF models of $13.33<R<14.26$km \cite{Reed:2021fv}. Using a range of Skyrme extended EDFs, \cite{Yue:2021ij} obtain $13.07<R<14.37$km from PREX data. It is notable that relatively high radii are obtained by these studies, which extrapolate the nuclear EOS to high densities. When we decouple the high-density EOS, allowing for EOSs that soften appreciably above 1.5$n_{\rm 0}$, one obtains lower radii when the PREX data is accounted for: from the PREX data alone we obtain $12.1<R<13.9$km, and combining that data with the NICER+LV data we obtain $11.8<R<13.0$km. This matches better the results of \cite{Tang:2021rt} $11.6<R<13$km who use 3 piece-wise polytropes, or a speed-of-sound, approach at high densities.

\cite{Zhang:2021ly} pointed out that the NICER data rule out a super-soft symmetry energy in the range 1-3$n_{0}$. This is born out in our work by the fact that the NICER+LV data significantly eliminates soft EOSs in that density range.

Moment of inertia predictions have been made in anticipation of a 10\% level measurement of the moment of inertia of pulsar B of the double pulsar J0737-7049 \cite{Carreau:2019qe,Greif:2020cr}, whose mass is 1.338$M_{\odot}$. From the NICER dataset, \cite{Silva:2021ys} obtain $1.68^{+0.53}_{-0.48}\times$10$^{45}$g cm$^{-2}$ and $1.64^{+0.52}_{-0.37}\times$10$^{45}$g cm$^{-2}$ compared with our $1.48^{+0.1}_{-0.14}\times$10$^{45}$g cm$^{-2}$ with the NICER+LV dataset. \cite{Miao:2021fj} obtain $1.27^{+0.18}_{-0.14}\times$10$^{45}$g cm$^{-2}$ or $1.29^{+0.25}_{-0.15}\times$10$^{45}$g cm$^{-2}$ depending on the EDF used. Our result is squarely in the range 1.3-1.6 $\times$10$^{45}$g cm$^{-2}$ from the study of nuclear equations of state from \cite{Worley:2008fr}. The maximum value of $I_{1.338}$ predicted is 1.9$\times$10$^{45}$g cm$^{-2}$ compared to the current 90\% upper limit from recent double pulsar timing measurements of \cite{Kramer:2021kq} of 3$\times$10$^{45}$g cm$^2$.

Our inferences of crust properties are the first to extract constraints from PREX from direct EDF calculations of the neutron skin of lead and combine them with astrophysical data using unified EOSs with high density polytropes. Our results are consistent with our previous work focusing solely on the crust EOS \cite{Balliet:2021lr,Newton:2021tg} and the similar calculations of \cite{Margueron:2018ab,Carreau:2019aa,Dinh-Thi:2021lr}. Our results for the ratio of the thickness and mass of pasta to that of the crust incorporating PREX with NICER+LV is $\Delta R_{\rm c}/\Delta R_{\rm p}=0.172^{+0.025}_{-0.124}$ and $\Delta M_{\rm c}/\Delta M_{\rm p}=0.51^{+0.08}_{-0.36}$, compared with $\Delta R_{\rm c}/\Delta R_{\rm p}=0.19^{+0.05}_{-0.07}$ and $\Delta M_{\rm c}/\Delta M_{\rm p}=0.57^{+0.10}_{-0.17}$ from PREX data alone \cite{Newton:2021tg}, and $\Delta R_{\rm c}/\Delta R_{\rm p}=0.128^{+0.047}_{-0.047}$ and $\Delta M_{\rm c}/\Delta M_{\rm p}=0.49^{+0.14}_{-0.14}$ from an analysis without PREX data, but incorporating information on the pure neutron matter EOS \cite{Dinh-Thi:2021lr}. 

For a $1.4$M$_{\odot}$ star, the relative moment of inertia fraction of the crust has a 68\% range from around 0.03 to 0.06 for the PREX data alone, 0.02 to 0.05 for the NICER+LV data alone, and and 0.04 to 0.07 for all the data combined. Previous systematic analyses of the moment of inertia fraction of the crust relative to the star obtained 95\% ranges of 0.02-0.06 (using PPs at high density) \cite{Steiner:2015kl} and 0.13-0.76 using meta-models. The constraint on $\Delta I_c / I$ from the Vela pulsar is $\Delta I_c / I > 0.016$ \cite{Link:1999rt} without taking into account entrainment of the crustal superfluid neutrons, and $\Delta I_c / I \gtrsim 0.08$ with entrainment \cite{Andersson:2012mz,Chamel:2013fr}. Our median values are all greater the lower limit of 0.016; indeed, the median values of the moment of inertia of the pasta phases are all greater than 0.016, highlighting the importance of understanding the effect of nuclear pasta on glitch mechanisms.

It is difficult to directly compare our results because we add a number of features that most studies do not: consistent crust EOSs - which add an additional physical requirement of a stable crust which filters a certain region of symmetry energy parameter space - and a large range of symmetry energy parameter space explored. We also, uniquely, parameterize our high density EOS by $M_{\rm max}$ and $I_{1.338}$ and our priors are uniform in those quantities rather than in the polytrope parameters. The choice of priors is know to be important, and a comparison with the more standard polytrope priors is underway. A thorough analysis of the difference between the various model inferences should be conducted.

\section{Conclusions}
\label{Conclusions}

Let us summarize our main findings:

\begin{enumerate}
    \item NICER+LV and PREX data are complementary - PREX data eliminates EOSs that are soft at low density, while NICER+LV data eliminate EOSs that are soft at higher densities.
    \item NICER+LV and PREX data are consistent with each other when it comes to bulk neutron star properties. While PREX predicts a high value of $L$ and NICER+LV predict lower values, they are compensated by the behavior of the high-density EOS. As pointed out in previous works, the PREX data requires softening of the EOS in the vicinity of saturation density.
    \item Astrophysical data provides more information about $K_{\rm sym}$ than neutron skin data. Both astrophysical and nuclear data provide information on $L$.
    \item Astrophysical data provides more information about the radius and moment of inertia of the star, while the PREX data provides more information about the gravitational binding energy of the star.
    \item Both astrophysical data and nuclear data provide information on the thickness, mass and moment of inertia of the crust and the pasta layers therein. The most powerful constraints are obtained by combining astrophysical and nuclear data.
\end{enumerate}

Our results are consistent with similar studies, although no previous study includes all of the features in the construction of EOS ensembles that we do here, and the origin of the differences in our predictions should be clarified.

We remind the reader that the results presented are not a statistically rigorous inference of neutron star properties. The data is applied as only a filter to the EOS distributions we have prepared, and a full Bayesian analysis is in progress. A full Bayesian approach would tend to increase the credible intervals, since it would now include the models in the tails of the distributions that have been cut-off here. Our intention here is to demonstrate the power of preparing ensembles of unified crust and core equations of state constructed with a full energy-density functional and appended by a sequence of high-density EOSs that allow us to get closer to exploring the full space of neutron star models. It is clear that powerful constraints on crust properties can be placed if we combine nuclear and astrophysical data in the way presented here. We do obtain for the first time ranges on the crust and pasta thickness, mass and moment of inertia incorporating electromagnetic probes of the neutron star radius, gravitational wave probes of the neutron star deformability and weak probes of the neutron skin of $^{208}$Pb, connected via a strong force model of the nuclear physics of nuclei and neutron stars. 

%

\emph{Acknowledgements}
This work is supported in part by the Physics and Astronomy Scholarship for Success (PASS) project funded by the NSF under grant No. 1643567 and the NASA grant 80NSSC18K1019.
\bibliographystyle{epj}
\bibliography{EOS_Paper}

\begin{thebibliography}{117}

\bibitem{Read:2009aa}
J.S. {Read}, B.D. {Lackey}, B.J. {Owen}, J.L. {Friedman}, Phys. Rev. D
  \textbf{79}, 124032 (2009), \texttt{0812.2163}

\bibitem{Ozel:2009kx}
F.~{{\"O}zel}, D.~{Psaltis}, Phys. Rev. D \textbf{80}, 103003 (2009),
  \texttt{0905.1959}

\bibitem{Steiner:2010aa}
A.W. {Steiner}, J.M. {Lattimer}, E.F. {Brown}, Astrophys. J. \textbf{722}, 33
  (2010), \texttt{1005.0811}

\bibitem{Abbott:2017aa}
B.P. {Abbott}, R.~{Abbott}, T.D. {Abbott}, F.~{Acernese}, K.~{Ackley},
  C.~{Adams}, T.~{Adams}, P.~{Addesso}, R.X. {Adhikari}, V.B. {Adya} et~al.,
  Astrophys. J.l \textbf{848}, L12 (2017), \texttt{1710.05833}

\bibitem{Abbott:2019kl}
B.P. {Abbott}, R.~{Abbott}, T.D. {Abbott}, F.~{Acernese}, K.~{Ackley},
  C.~{Adams}, T.~{Adams}, P.~{Addesso}, R.X. {Adhikari}, V.B. {Adya} et~al.,
  Physical Review X \textbf{9}, 011001 (2019), \texttt{1805.11579}

\bibitem{Riley:2019aa}
T.E. {Riley}, A.L. {Watts}, S.~{Bogdanov}, P.S. {Ray}, R.M. {Ludlam},
  S.~{Guillot}, Z.~{Arzoumanian}, C.L. {Baker}, A.V. {Bilous}, D.~{Chakrabarty}
  et~al., Astrophys. J.l \textbf{887}, L21 (2019), \texttt{1912.05702}

\bibitem{Raaijmakers:2019sf}
G.~{Raaijmakers}, T.E. {Riley}, A.L. {Watts}, S.K. {Greif}, S.M. {Morsink},
  K.~{Hebeler}, A.~{Schwenk}, T.~{Hinderer}, S.~{Nissanke}, S.~{Guillot}
  et~al., Astrophys. J.l \textbf{887}, L22 (2019), \texttt{1912.05703}

\bibitem{Miller:2019jt}
M.C. {Miller}, F.K. {Lamb}, A.J. {Dittmann}, S.~{Bogdanov}, Z.~{Arzoumanian},
  K.C. {Gendreau}, S.~{Guillot}, A.K. {Harding}, W.C.G. {Ho}, J.M. {Lattimer}
  et~al., Astrophys. J.l \textbf{887}, L24 (2019), \texttt{1912.05705}

\bibitem{Miller:2021hb}
M.C. {Miller}, F.K. {Lamb}, A.J. {Dittmann}, S.~{Bogdanov}, Z.~{Arzoumanian},
  K.C. {Gendreau}, S.~{Guillot}, W.C.G. {Ho}, J.M. {Lattimer}, M.~{Loewenstein}
  et~al., arXiv e-prints arXiv:2105.06979 (2021), \texttt{2105.06979}

\bibitem{Riley:2021sp}
T.E. {Riley}, A.L. {Watts}, P.S. {Ray}, S.~{Bogdanov}, S.~{Guillot}, S.M.
  {Morsink}, A.V. {Bilous}, Z.~{Arzoumanian}, D.~{Choudhury}, J.S. {Deneva}
  et~al., arXiv e-prints arXiv:2105.06980 (2021), \texttt{2105.06980}

\bibitem{Raaijmakers:2021hc}
G.~{Raaijmakers}, S.K. {Greif}, K.~{Hebeler}, T.~{Hinderer}, S.~{Nissanke},
  A.~{Schwenk}, T.E. {Riley}, A.L. {Watts}, J.M. {Lattimer}, W.C.G. {Ho}, arXiv
  e-prints arXiv:2105.06981 (2021), \texttt{2105.06981}

\bibitem{Raithel:2018ai}
C.A. {Raithel}, F.~{{\"O}zel}, D.~{Psaltis}, Astrophys. J.l \textbf{857}, L23
  (2018), \texttt{1803.07687}

\bibitem{Greif:2020cr}
S.K. {Greif}, K.~{Hebeler}, J.M. {Lattimer}, C.J. {Pethick}, A.~{Schwenk},
  Astrophys. J. \textbf{901}, 155 (2020), \texttt{2005.14164}

\bibitem{Al-Mamun:2021ue}
M.~{Al-Mamun}, A.W. {Steiner}, J.~{N{\"a}ttil{\"a}}, J.~{Lange},
  R.~{O'Shaughnessy}, I.~{Tews}, S.~{Gandolfi}, C.~{Heinke}, S.~{Han}, Phys.
  Rev. Lett. \textbf{126}, 061101 (2021), \texttt{2008.12817}

\bibitem{Tews:2018fj}
I.~{Tews}, J.~{Carlson}, S.~{Gandolfi}, S.~{Reddy}, Astrophys. J. \textbf{860},
  149 (2018), \texttt{1801.01923}

\bibitem{Annala:2020wd}
E.~{Annala}, T.~{Gorda}, A.~{Kurkela}, J.~{N{\"a}ttil{\"a}}, A.~{Vuorinen},
  Nature Physics \textbf{16}, 907 (2020), \texttt{1903.09121}

\bibitem{Essick:2021vn}
R.~{Essick}, P.~{Landry}, A.~{Schwenk}, I.~{Tews}, arXiv e-prints
  arXiv:2107.05528 (2021), \texttt{2107.05528}

\bibitem{Lindblom:2010vf}
L.~{Lindblom}, Phys. Rev. D \textbf{82}, 103011 (2010), \texttt{1009.0738}

\bibitem{Landry:2019te}
P.~{Landry}, R.~{Essick}, Phys. Rev. D \textbf{99}, 084049 (2019),
  \texttt{1811.12529}

\bibitem{Landry:2020wv}
P.~{Landry}, R.~{Essick}, K.~{Chatziioannou}, Phys. Rev. D \textbf{101}, 123007
  (2020), \texttt{2003.04880}

\bibitem{Essick:2020uy}
R.~{Essick}, P.~{Landry}, D.E. {Holz}, Phys. Rev. D \textbf{101}, 063007
  (2020), \texttt{1910.09740}

\bibitem{Legred:2021wn}
I.~{Legred}, K.~{Chatziioannou}, R.~{Essick}, S.~{Han}, P.~{Landry}, arXiv
  e-prints arXiv:2106.05313 (2021), \texttt{2106.05313}

\bibitem{Ferreira:2019wh}
M.~{Ferreira}, C.~{Provid{\^e}ncia}, arXiv e-prints arXiv:1910.05554 (2019),
  \texttt{1910.05554}

\bibitem{Abrahamyan:2012ai}
S.~{Abrahamyan}, Z.~{Ahmed}, H.~{Albataineh}, K.~{Aniol}, D.S. {Armstrong},
  W.~{Armstrong}, T.~{Averett}, B.~{Babineau}, A.~{Barbieri}, V.~{Bellini}
  et~al., Phys. Rev. Lett. \textbf{108}, 112502 (2012), \texttt{1201.2568}

\bibitem{Becker:2018aa}
D.~{Becker}, R.~{Bucoveanu}, C.~{Grzesik}, K.~{Imai}, R.~{Kempf}, M.~{Molitor},
  A.~{Tyukin}, M.~{Zimmermann}, D.~{Armstrong}, K.~{Aulenbacher} et~al.,
  European Physical Journal A \textbf{54}, 208 (2018), \texttt{1802.04759}

\bibitem{Thiel:2019aa}
M.~{Thiel}, C.~{Sfienti}, J.~{Piekarewicz}, C.J. {Horowitz},
  M.~{Vanderhaeghen}, Journal of Physics G Nuclear Physics \textbf{46}, 093003
  (2019), \texttt{1904.12269}

\bibitem{Adhikari:2021tm}
D.~{Adhikari}, H.~{Albataineh}, D.~{Androic}, K.~{Aniol}, D.S. {Armstrong},
  T.~{Averett}, C.~{Ayerbe Gayoso}, S.~{Barcus}, V.~{Bellini}, R.S.
  {Beminiwattha} et~al., Phys. Rev. Lett. \textbf{126}, 172502 (2021),
  \texttt{2102.10767}

\bibitem{Tamii:2011lr}
A.~{Tamii}, I.~{Poltoratska}, P.~{von Neumann-Cosel}, Y.~{Fujita}, T.~{Adachi},
  C.A. {Bertulani}, J.~{Carter}, M.~{Dozono}, H.~{Fujita}, K.~{Fujita} et~al.,
  Phys. Rev. Lett. \textbf{107}, 062502 (2011), \texttt{1104.5431}

\bibitem{Hashimoto:2015fj}
T.~{Hashimoto}, A.M. {Krumbholz}, P.G. {Reinhard}, A.~{Tamii}, P.~{von
  Neumann-Cosel}, T.~{Adachi}, N.~{Aoi}, C.A. {Bertulani}, H.~{Fujita},
  Y.~{Fujita} et~al., Phys. Rev. C \textbf{92}, 031305 (2015),
  \texttt{1503.08321}

\bibitem{von-Neumann-Cosel:2015mz}
P.~{von Neumann-Cosel}, \emph{{Complete electric dipole response in $^{120}$Sn
  and $^{208}$Pb and implications for neutron skin and symmetry energy}}, in
  \emph{Journal of Physics Conference Series} (2015), Vol. 590 of \emph{Journal
  of Physics Conference Series}, p. 012017

\bibitem{Birkhan:2017fk}
J.~{Birkhan}, M.~{Miorelli}, S.~{Bacca}, S.~{Bassauer}, C.A. {Bertulani},
  G.~{Hagen}, H.~{Matsubara}, P.~{von Neumann-Cosel}, T.~{Papenbrock},
  N.~{Pietralla} et~al., Phys. Rev. Lett. \textbf{118}, 252501 (2017),
  \texttt{1611.07072}

\bibitem{Gandolfi:2015nr}
S.~{Gandolfi}, A.~{Gezerlis}, J.~{Carlson}, Annual Review of Nuclear and
  Particle Science \textbf{65}, 303 (2015), \texttt{1501.05675}

\bibitem{Hagen:2016ul}
G.~{Hagen}, A.~{Ekstr{\"o}m}, C.~{Forss{\'e}n}, G.R. {Jansen}, W.~{Nazarewicz},
  T.~{Papenbrock}, K.A. {Wendt}, S.~{Bacca}, N.~{Barnea}, B.~{Carlsson} et~al.,
  Nature Physics \textbf{12}, 186 (2016)

\bibitem{Tews:2016ty}
I.~{Tews}, S.~{Gandolfi}, A.~{Gezerlis}, A.~{Schwenk}, Phys. Rev. C
  \textbf{93}, 024305 (2016), \texttt{1507.05561}

\bibitem{Drischler:2020aa}
C.~{Drischler}, J.A. {Melendez}, R.J. {Furnstahl}, D.R. {Phillips}, arXiv
  e-prints arXiv:2004.07805 (2020), \texttt{2004.07805}

\bibitem{Drischler:2020ab}
C.~{Drischler}, R.J. {Furnstahl}, J.A. {Melendez}, D.R. {Phillips}, arXiv
  e-prints arXiv:2004.07232 (2020), \texttt{2004.07232}

\bibitem{Hebeler:2010rw}
K.~{Hebeler}, J.M. {Lattimer}, C.J. {Pethick}, A.~{Schwenk}, Phys. Rev. Lett.
  \textbf{105}, 161102 (2010), \texttt{1007.1746}

\bibitem{Gandolfi:2010jk}
S.~{Gandolfi}, A.Y. {Illarionov}, S.~{Fantoni}, J.C. {Miller}, F.~{Pederiva},
  K.E. {Schmidt}, Mon. Not. R. Astron. Soc \textbf{404}, L35 (2010),
  \texttt{0909.3487}

\bibitem{Pang:2021gf}
P.T.H. {Pang}, I.~{Tews}, M.W. {Coughlin}, M.~{Bulla}, C.~{Van Den Broeck},
  T.~{Dietrich}, arXiv e-prints arXiv:2105.08688 (2021), \texttt{2105.08688}

\bibitem{Huth:2021wr}
S.~{Huth}, C.~{Wellenhofer}, A.~{Schwenk}, Phys. Rev. C \textbf{103}, 025803
  (2021), \texttt{2009.08885}

\bibitem{Huth:2021lr}
S.~{Huth}, P.T.H. {Pang}, I.~{Tews}, T.~{Dietrich}, A.~{Le F{\`e}vre},
  A.~{Schwenk}, W.~{Trautmann}, K.~{Agarwal}, M.~{Bulla}, M.W. {Coughlin}
  et~al., arXiv e-prints arXiv:2107.06229 (2021), \texttt{2107.06229}

\bibitem{Brown:2000aa}
B.A. {Brown}, Phys. Rev. Lett. \textbf{85}, 5296 (2000)

\bibitem{Horowitz:2001aa}
C.J. {Horowitz}, J.~{Piekarewicz}, Phys. Rev. Lett. \textbf{86}, 5647 (2001),
  \texttt{astro-ph/0010227}

\bibitem{Steiner:2008aa}
A.W. {Steiner}, B.A. {Li}, M.~{Prakash}, \emph{{Ramifications of the Nuclear
  Symmetry Energy for Neutron Stars, Nuclei and Heavy-Ion Collisions}}, in
  \emph{Exotic States of Nuclear Matter} (2008), pp. 47--54, \texttt{0711.4652}

\bibitem{Fattoyev:2018aa}
F.J. {Fattoyev}, J.~{Piekarewicz}, C.J. {Horowitz}, Phys. Rev. Lett.
  \textbf{120}, 172702 (2018), \texttt{1711.06615}

\bibitem{Margueron:2018aa}
J.~{Margueron}, R.~{Hoffmann Casali}, F.~{Gulminelli}, Phys. Rev. C
  \textbf{97}, 025805 (2018), \texttt{1708.06894}

\bibitem{Margueron:2018ab}
J.~{Margueron}, R.~{Hoffmann Casali}, F.~{Gulminelli}, Phys. Rev. C
  \textbf{97}, 025806 (2018), \texttt{1708.06895}

\bibitem{Lim:2019xr}
Y.~{Lim}, J.W. {Holt}, European Physical Journal A \textbf{55}, 209 (2019),
  \texttt{1902.05502}

\bibitem{Zhang:2021ly}
N.B. {Zhang}, B.A. {Li}, arXiv e-prints arXiv:2105.11031 (2021),
  \texttt{2105.11031}

\bibitem{Biswas:2021bh}
B.~{Biswas}, P.~{Char}, R.~{Nandi}, S.~{Bose}, Phys. Rev. D \textbf{103},
  103015 (2021), \texttt{2008.01582}

\bibitem{Lee:2021eu}
H.K. {Lee}, Y.L. {Ma}, W.G. {Paeng}, M.~{Rho}, arXiv e-prints arXiv:2107.01879
  (2021), \texttt{2107.01879}

\bibitem{Tang:2021rt}
S.P. {Tang}, J.L. {Jiang}, M.Z. {Han}, Y.Z. {Fan}, D.M. {Wei}, Phys. Rev. D
  \textbf{104}, 063032 (2021), \texttt{2106.04204}

\bibitem{Tsang:2012qy}
M.B. {Tsang}, J.R. {Stone}, F.~{Camera}, P.~{Danielewicz}, S.~{Gandolfi},
  K.~{Hebeler}, C.J. {Horowitz}, J.~{Lee}, W.G. {Lynch}, Z.~{Kohley} et~al.,
  Phys. Rev. C \textbf{86}, 015803 (2012), \texttt{1204.0466}

\bibitem{Li:2014fj}
B.A. {Li}, {\`A}.~{Ramos}, G.~{Verde}, I.~{Vida{\~n}a}, European Physical
  Journal A \textbf{50}, 9 (2014)

\bibitem{Lattimer:2014uq}
J.M. {Lattimer}, A.W. {Steiner}, European Physical Journal A \textbf{50}, 40
  (2014), \texttt{1403.1186}

\bibitem{Horowitz:2014aa}
C.J. {Horowitz}, E.F. {Brown}, Y.~{Kim}, W.G. {Lynch}, R.~{Michaels}, A.~{Ono},
  J.~{Piekarewicz}, M.B. {Tsang}, H.H. {Wolter}, Journal of Physics G Nuclear
  Physics \textbf{41}, 093001 (2014), \texttt{1401.5839}

\bibitem{Tsang:2019ab}
M.B. {Tsang}, W.G. {Lynch}, P.~{Danielewicz}, C.Y. {Tsang}, Physics Letters B
  \textbf{795}, 533 (2019), \texttt{1906.02180}

\bibitem{Lynch:2021qp}
W.G. {Lynch}, M.B. {Tsang}, arXiv e-prints arXiv:2106.10119 (2021),
  \texttt{2106.10119}

\bibitem{Roca-Maza:2011aa}
X.~{Roca-Maza}, M.~{Centelles}, X.~{Vi{\~n}as}, M.~{Warda}, Phys. Rev. Lett.
  \textbf{106}, 252501 (2011), \texttt{1103.1762}

\bibitem{Biswas:2021pd}
B.~{Biswas}, arXiv e-prints arXiv:2105.02886 (2021), \texttt{2105.02886}

\bibitem{Du:2021fk}
X.~{Du}, A.W. {Steiner}, J.W. {Holt}, arXiv e-prints arXiv:2107.06697 (2021),
  \texttt{2107.06697}

\bibitem{Yue:2021ij}
T.G. {Yue}, L.W. {Chen}, Z.~{Zhang}, Y.~{Zhou}, arXiv e-prints arXiv:2102.05267
  (2021), \texttt{2102.05267}

\bibitem{Sellahewa:2014xy}
R.~{Sellahewa}, A.~{Rios}, Phys. Rev. C \textbf{90}, 054327 (2014),
  \texttt{1407.8138}

\bibitem{Vinas:2021fv}
X.~{Vi{\~n}as}, C.~{Gonzalez-Boquera}, M.~{Centelles}, C.~{Mondal}, L.M.
  {Robledo}, Symmetry \textbf{13}, 1613 (2021), \texttt{2109.02520}

\bibitem{Dutra:2014fy}
M.~{Dutra}, O.~{Louren{\c{c}}o}, S.S. {Avancini}, B.V. {Carlson}, A.~{Delfino},
  D.P. {Menezes}, C.~{Provid{\^e}ncia}, S.~{Typel}, J.R. {Stone}, Phys. Rev. C
  \textbf{90}, 055203 (2014), \texttt{1405.3633}

\bibitem{Beloin:2019oq}
S.~{Beloin}, S.~{Han}, A.W. {Steiner}, K.~{Odbadrakh}, Phys. Rev. C
  \textbf{100}, 055801 (2019), \texttt{1812.00494}

\bibitem{Tsang:2018yq}
M.B. {Tsang}, C.Y. {Tsang}, P.~{Danielewicz}, W.G. {Lynch}, F.J. {Fattoyev},
  arXiv e-prints arXiv:1811.04888 (2018), \texttt{1811.04888}

\bibitem{Miao:2021fj}
Z.~{Miao}, A.~{Li}, arXiv e-prints arXiv:2107.07979 (2021), \texttt{2107.07979}

\bibitem{Li:2021kx}
A.~{Li}, Z.~{Miao}, S.~{Han}, B.~{Zhang}, Astrophys. J. \textbf{913}, 27
  (2021), \texttt{2103.15119}

\bibitem{Chen:2009hh}
L.W. {Chen}, B.J. {Cai}, C.M. {Ko}, B.A. {Li}, C.~{Shen}, J.~{Xu}, Phys. Rev. C
  \textbf{80}, 014322 (2009), \texttt{0905.4323}

\bibitem{Centelles:2009aa}
M.~{Centelles}, X.~{Roca-Maza}, X.~{Vi{\~n}as}, M.~{Warda}, Phys. Rev. Lett.
  \textbf{102}, 122502 (2009), \texttt{0806.2886}

\bibitem{Raduta:2018aa}
A.R. {Raduta}, F.~{Gulminelli}, Phys. Rev. C \textbf{97}, 064309 (2018),
  \texttt{1712.05973}

\bibitem{Newton:2021dq}
W.G. {Newton}, G.~{Crocombe}, Phys. Rev. C \textbf{103}, 064323 (2021),
  \texttt{2008.00042}

\bibitem{Carreau:2019aa}
T.~{Carreau}, F.~{Gulminelli}, J.~{Margueron}, European Physical Journal A
  \textbf{55}, 188 (2019), \texttt{1902.07032}

\bibitem{Dinh-Thi:2021lr}
H.~{Dinh Thi}, T.~{Carreau}, A.F. {Fantina}, F.~{Gulminelli}, Astron.
  Astrophys. \textbf{654}, A114 (2021), \texttt{2109.13638}

\bibitem{Balliet:2021lr}
L.E. {Balliet}, W.G. {Newton}, S.~{Cantu}, S.~{Budimir}, Astrophys. J.
  \textbf{918}, 79 (2021), \texttt{2009.07696}

\bibitem{Newton:2021tg}
W.G. {Newton}, R.~{Preston}, L.~{Balliet}, M.~{Ross}, arXiv e-prints
  arXiv:2111.07969 (2021), \texttt{2111.07969}

\bibitem{Gamba:2020ul}
R.~{Gamba}, J.S. {Read}, L.E. {Wade}, Classical and Quantum Gravity
  \textbf{37}, 025008 (2020), \texttt{1902.04616}

\bibitem{Suleiman:2021ve}
L.~{Suleiman}, M.~{Fortin}, J.L. {Zdunik}, P.~{Haensel}, Phys. Rev. C
  \textbf{104}, 015801 (2021), \texttt{2106.12845}

\bibitem{Douchin:2001bq}
F.~{Douchin}, P.~{Haensel}, Astron. Astrophys. \textbf{380}, 151 (2001),
  \texttt{astro-ph/0111092}

\bibitem{Fortin:2016uq}
M.~{Fortin}, C.~{Provid{\^e}ncia}, A.R. {Raduta}, F.~{Gulminelli}, J.L.
  {Zdunik}, P.~{Haensel}, M.~{Bejger}, Phys. Rev. C \textbf{94}, 035804 (2016),
  \texttt{1604.01944}

\bibitem{McNeil-Forbes:2019wx}
M.~{McNeil Forbes}, S.~{Bose}, S.~{Reddy}, D.~{Zhou}, A.~{Mukherjee}, S.~{De},
  arXiv e-prints arXiv:1904.04233 (2019), \texttt{1904.04233}

\bibitem{Steiner:2009wo}
A.W. {Steiner}, A.L. {Watts}, Physical Review Letters \textbf{103}, 181101
  (2009), \texttt{0902.1683}

\bibitem{Gearheart:2011tg}
M.~{Gearheart}, W.G. {Newton}, J.~{Hooker}, B.A. {Li}, Mon. Not. R. Astron. Soc
  \textbf{418}, 2343 (2011), \texttt{1106.4875}

\bibitem{Sotani:2012oz}
H.~{Sotani}, K.~{Nakazato}, K.~{Iida}, K.~{Oyamatsu}, Physical Review Letters
  \textbf{108}, 201101 (2012), \texttt{1202.6242}

\bibitem{Brown:2009aa}
E.F. {Brown}, A.~{Cumming}, Astrophys. J. \textbf{698}, 1020 (2009),
  \texttt{0901.3115}

\bibitem{Horowitz:2015rt}
C.J. {Horowitz}, D.K. {Berry}, C.M. {Briggs}, M.E. {Caplan}, A.~{Cumming}, A.S.
  {Schneider}, Physical Review Letters \textbf{114}, 031102 (2015),
  \texttt{1410.2197}

\bibitem{Newton:2013dz}
W.G. {Newton}, K.~{Murphy}, J.~{Hooker}, B.A. {Li}, Astrophys. J.l
  \textbf{779}, L4 (2013), \texttt{1308.2137}

\bibitem{Tsang:2012aa}
D.~{Tsang}, J.S. {Read}, T.~{Hinderer}, A.L. {Piro}, R.~{Bondarescu}, Phys.
  Rev. Lett. \textbf{108}, 011102 (2012), \texttt{1110.0467}

\bibitem{Neill:2021lq}
D.~{Neill}, W.G. {Newton}, D.~{Tsang}, Mon. Not. R. Astron. Soc \textbf{504},
  1129 (2021), \texttt{2012.10322}

\bibitem{Pons:2013ly}
J.A. {Pons}, D.~{Vigan{\`o}}, N.~{Rea}, Nature Physics \textbf{9}, 431 (2013),
  \texttt{1304.6546}

\bibitem{Wen:2012aa}
D.H. {Wen}, W.G. {Newton}, B.A. {Li}, Phys. Rev. C \textbf{85}, 025801 (2012),
  \texttt{1110.5985}

\bibitem{Vidana:2012aa}
I.~{Vida{\~n}a}, Phys. Rev. C \textbf{85}, 045808 (2012), \texttt{1202.4731}

\bibitem{Lim:2017aa}
Y.~{Lim}, J.W. {Holt}, Phys. Rev. C \textbf{95}, 065805 (2017)

\bibitem{Kramer:2021kq}
M.~{Kramer}, I.H. {Stairs}, R.N. {Manchester}, N.~{Wex}, A.T. {Deller}, W.A.
  {Coles}, M.~{Ali}, M.~{Burgay}, F.~{Camilo}, I.~{Cognard} et~al., Physical
  Review X \textbf{11}, 041050 (2021), \texttt{2112.06795}

\bibitem{Newton:2013sp}
W.G. {Newton}, M.~{Gearheart}, B.A. {Li}, Astrophys. J.s \textbf{204}, 9
  (2013), \texttt{1110.4043}

\bibitem{Capano:2020ai}
C.D. {Capano}, I.~{Tews}, S.M. {Brown}, B.~{Margalit}, S.~{De}, S.~{Kumar},
  D.A. {Brown}, B.~{Krishnan}, S.~{Reddy}, Nature Astronomy \textbf{4}, 625
  (2020), \texttt{1908.10352}

\bibitem{Tews:2019ff}
I.~{Tews}, J.~{Margueron}, S.~{Reddy}, European Physical Journal A \textbf{55},
  97 (2019), \texttt{1901.09874}

\bibitem{Yagi:2013aa}
K.~{Yagi}, N.~{Yunes}, Phys. Rev. D \textbf{88}, 023009 (2013),
  \texttt{1303.1528}

\bibitem{Yagi:2013ab}
K.~{Yagi}, N.~{Yunes}, Science \textbf{341}, 365 (2013), \texttt{1302.4499}

\bibitem{Yagi:2017aa}
K.~{Yagi}, N.~{Yunes}, Physics Reports \textbf{681}, 1 (2017),
  \texttt{1608.02582}

\bibitem{Demorest2010a-mass}
P.B. {Demorest}, T.~{Pennucci}, S.M. {Ransom}, M.S.E. {Roberts}, J.W.T.
  {Hessels}, Nature \textbf{467}, 1081 (2010), \texttt{1010.5788}

\bibitem{Antoniadis2013a-mass}
J.~{Antoniadis}, P.C.C. {Freire}, N.~{Wex}, T.M. {Tauris}, R.S. {Lynch}, M.H.
  {van Kerkwijk}, M.~{Kramer}, C.~{Bassa}, V.S. {Dhillon}, T.~{Driebe} et~al.,
  Science \textbf{340}, 448 (2013), \texttt{1304.6875}

\bibitem{Shibata:2017aa}
M.~{Shibata}, S.~{Fujibayashi}, K.~{Hotokezaka}, K.~{Kiuchi}, K.~{Kyutoku},
  Y.~{Sekiguchi}, M.~{Tanaka}, Phys. Rev. D \textbf{96}, 123012 (2017),
  \texttt{1710.07579}

\bibitem{Margalit:2017aa}
B.~{Margalit}, B.D. {Metzger}, Astrophys. J.l \textbf{850}, L19 (2017),
  \texttt{1710.05938}

\bibitem{Ruiz:2018aa}
M.~{Ruiz}, S.L. {Shapiro}, A.~{Tsokaros}, Phys. Rev. D \textbf{97}, 021501
  (2018), \texttt{1711.00473}

\bibitem{Rezzolla:2018aa}
L.~{Rezzolla}, E.R. {Most}, L.R. {Weih}, Astrophys. J.l \textbf{852}, L25
  (2018), \texttt{1711.00314}

\bibitem{Abbott:2018fe}
B.P. {Abbott}, R.~{Abbott}, T.D. {Abbott}, F.~{Acernese}, K.~{Ackley},
  C.~{Adams}, T.~{Adams}, P.~{Addesso}, R.X. {Adhikari}, V.B. {Adya} et~al.,
  Phys. Rev. Lett. \textbf{121}, 161101 (2018), \texttt{1805.11581}

\bibitem{Radice:2019xy}
D.~{Radice}, L.~{Dai}, European Physical Journal A \textbf{55}, 50 (2019),
  \texttt{1810.12917}

\bibitem{Reed:2021fv}
B.T. {Reed}, F.J. {Fattoyev}, C.J. {Horowitz}, J.~{Piekarewicz}, Phys. Rev.
  Lett. \textbf{126}, 172503 (2021), \texttt{2101.03193}

\bibitem{Carreau:2019qe}
T.~{Carreau}, F.~{Gulminelli}, J.~{Margueron}, Phys. Rev. C \textbf{100},
  055803 (2019), \texttt{1810.00719}

\bibitem{Silva:2021ys}
H.O. {Silva}, A.M. {Holgado}, A.~{C{\'a}rdenas-Avenda{\~n}o}, N.~{Yunes}, Phys.
  Rev. Lett. \textbf{126}, 181101 (2021), \texttt{2004.01253}

\bibitem{Worley:2008fr}
A.~{Worley}, P.G. {Krastev}, B.A. {Li}, arXiv e-prints arXiv:0801.1653 (2008),
  \texttt{0801.1653}

\bibitem{Steiner:2015kl}
A.W. {Steiner}, S.~{Gandolfi}, F.J. {Fattoyev}, W.G. {Newton}, Phys. Rev. C
  \textbf{91}, 015804 (2015), \texttt{1403.7546}

\bibitem{Link:1999rt}
B.~{Link}, R.I. {Epstein}, J.M. {Lattimer}, Phys. Rev. Lett. \textbf{83}, 3362
  (1999), \texttt{astro-ph/9909146}

\bibitem{Andersson:2012mz}
N.~{Andersson}, K.~{Glampedakis}, W.C.G. {Ho}, C.M. {Espinoza}, Phys. Rev.
  Lett. \textbf{109}, 241103 (2012), \texttt{1207.0633}

\bibitem{Chamel:2013fr}
N.~{Chamel}, Phys. Rev. Lett. \textbf{110}, 011101 (2013), \texttt{1210.8177}

\end{thebibliography}
%
%
%
%

\end{document}